\documentclass[acmsmall]{acmart}

\AtBeginDocument{%
  \providecommand\BibTeX{{%
    \normalfont B\kern-0.5em{\scshape i\kern-0.25em b}\kern-0.8em\TeX}}}

\acmJournal{TODAES}

\usepackage[utf8]{inputenc}
\usepackage{caption}
\usepackage[perpage,symbol]{footmisc}
\usepackage[english]{babel}
\usepackage{multirow,multicol}
\usepackage{amsmath, bm} 
\usepackage{pbox, booktabs, rotating}
\usepackage{subfigure}
\usepackage{array, makecell, lipsum}
\usepackage{enumerate}
\usepackage{threeparttable}
\usepackage{multirow,hyphenat,balance}
\usepackage{placeins}
\usepackage{float,color}
\usepackage{ragged2e}
\usepackage{enumerate}
\usepackage{etoolbox}
\usepackage[linesnumbered,ruled]{algorithm2e}
\usepackage{paralist}
\usepackage{cases}
\usepackage{url}
\usepackage{framed}
\usepackage{flushend}
\newcolumntype{M}[1]{>{\centering\arraybackslash}m{#1}}

\newcommand*{\eg}{{e.g.}\@\xspace}
\newcommand*{\el}{et al.\@\xspace}

\begin{document}

\markboth{}{Jain et al.}

\title{TAAL: Tampering Attack on Any Key-based Logic Locked Circuits}

%\author{Ayush Jain, \IEEEmembership{Student Member, IEEE}, Ziqi Zhou, \IEEEmembership{Student Member, IEEE} and Ujjwal Guin, \IEEEmembership{Member, IEEE}
%\thanks{Ayush Jain, Ziqi Zhou and Ujjwal Guin are with the Department of Electrical and Computer Engineering, Auburn University, AL, USA (e-mail: \{ayush.jain, ziqi.zhou and ujjwal.guin\}@auburn.edu).\ }
%}

\author{Ayush Jain} \email{ayush.jain@auburn.edu}
\author{Ziqi Zhou} \email{ziqi.zhou@auburn.edu}
\author{Ujjwal Guin} \email{ujjwal.guin@auburn.edu}
\affiliation{
\institution{Auburn University}
\department{Department of Electrical and Computer Engineering}
\city{Auburn}
\state{AL}
\country{USA}}

%\maketitle

% \author{Ayush Jain\email{1} \and Ziqi Zhou\email{2} \and Ujjwal Guin\email{3}}
% \institute{Department of Electrical and Computer Engineering, Auburn University, USA 

% \email{{azj0041,zhouziq,ujjwal.guin}@auburn.edu}}

%\maketitle
%\keywords[Logic locking, hardware Trojans]{Logic locking \and IP Piracy \and IC Overproduction \and Hardware Trojans \and Tampering}

\begin{abstract}
Due to the globalization of semiconductor manufacturing and test processes, the system-on-a-chip (SoC) designers no longer design the complete SoC and manufacture chips on their own. This outsourcing of the design and manufacturing of Integrated Circuits (ICs) has resulted in several threats, such as overproduction of ICs, sale of out-of-specification/rejected ICs, and piracy of Intellectual Properties (IPs). Logic locking has emerged as a promising defense strategy against these threats. However, various attacks about the extraction of secret keys have undermined the security of logic locking techniques. Over the years, researchers have proposed different techniques to prevent existing attacks. In this paper, we propose a novel attack that can break any logic locking techniques that rely on the stored secret key. This proposed \textit{TAAL} attack is based on implanting a hardware Trojan in the netlist, which leaks the secret key to an adversary once activated. As an untrusted foundry can extract the netlist of a design from the layout/mask information, it is feasible to implement such a hardware Trojan. All three proposed types of \textit{TAAL} attacks can be used for extracting secret keys. We have introduced the models for both the combinational and sequential hardware Trojans that evade manufacturing tests. An adversary only needs to choose one hardware Trojan out of a large set of all possible Trojans to launch the \textit{TAAL} attack. 
\end{abstract}

\vspace{10px}


\maketitle
\renewcommand{\shortauthors}{Jain et al.}

\section{Introduction}
The continuous addition of new functionality in a system-on-a-chip (SoC) has enforced designers to adopt newer and lower technology nodes to manufacture chips primarily to reduce the overall area and the resultant cost of a chip. Building and maintaining such a fabrication plant (foundry) requires a multi-billion dollar investment~\cite{YehFabCost2012}. As a result, the semiconductor industry has moved towards horizontal integration, where an SoC designer acquires intellectual properties (IPs) from many different vendors and sends the design to a foundry for manufacturing, which is generally located offshore.       

The hardware layers that were assumed to be trusted are no longer true with the outsourcing of IC fabrication in a globalized and distributed design flow, including multiple entities. Third-party IPs, fabrication, and test facilities of chips represent security threats to the current horizontal integration of the production. The security threats posed by these entities include -- $(i)$ overproduction of ICs~\cite{roy2008epic, alkabani2007active, chakraborty2008hardware, alkabani2007remote, huang2008ic, baumgarten2010preventing, guin2016fortis}, where an untrusted foundry fabricates more chips without the consent of the SoC designer in order to generate revenue by selling them in the market, $(ii)$ sale of out-of-specification/rejected ICs~\cite{contreras2013secure, rahman2014csst, guin2016fortis, GuinVTS2017, zhang2018chip}, and $(iii)$ IP piracy~\cite{castillo2007ipp, tehranipoor2011introduction, tehranipoor2015counterfeit, bhunia2018hardware, zhang2018chip}, where an entity in the supply chain can use, modify and/or sell functional IPs illegally. Over the years, researchers have proposed different techniques to prevent the aforementioned attacks and they are IC metering~\cite{roy2008epic, alkabani2007active, alkabani2007remote, koushanfar2001hardware}, logic locking~\cite{roy2008epic, rajendran2012security, guin2016fortis}, hardware watermarking~\cite{charbon1998hierarchical, kahng2001constraint, qu2007intellectual}, and split manufacturing~\cite{jarvis2007split, vaidyanathan2014efficient}.    

Logic locking has emerged as a promising technique and gained significant attention from the researchers to address the different threats emerging from untrusted manufacturing. In logic locking, the netlist of a circuit is locked so that it produces incorrect results unless it is programmed with a secret key. The locks are generally inserted in the netlist using XOR gates. Traditional logic locking methods involved selection of key gate location as random selection~\cite{roy2008epic, GuinVTS2017, GuinTVLSI2018}, strong interference-based selection~\cite{rajendran2012security}, and fault-analysis based selection~\cite{rajendran2015fault}. Over the years, researchers have also proposed different attacks to extract the secret key and undermine the locking mechanisms. Boolean Satisfiability (SAT)-based attacks have demonstrated effective ways of extracting the secret keys. Countermeasures are also proposed so that SAT-based attacks become infeasible. 

This paper shows that any locked circuit, where the secret key is stored in an on-chip memory no matter whether it is tamper-proof or not, can be broken by inserting a hardware Trojan. We present this attack as \textbf{TAAL}: \textbf{T}ampering \textbf{A}ttack on \textbf{A}ny \textbf{L}ocked circuit that uses a stored key for locking the netlist. This attack can defeat the security measures provided from any existing logic locking methods. \textit{We believe that we are the first to show that any key-based logic locked circuit can be exploited by inserting a hardware Trojan in the netlist}. The contributions of this paper are described as follows:

\begin{itemize}
    \item We propose a novel attack based on malicious modifications of the netlist to target any key-based locked circuit. The attacking approach is to tamper the locked netlist in order to extract the secret key information. Once the valid key information is extracted from an activated IC, an untrusted foundry can unlock any number of chips and sell overproduced and defective chips. As this attack applies to any key-based locked circuits, an adversary can undermine any secure solutions proposed so far to prevent overproduction, sourcing defective/out-of-spec chips, and IP piracy. Note that different researchers have proposed to use logic locking to prevent hardware Trojans~\cite{yasin2015improving, yasin2015transforming, liu2014embedded, yu2017exploiting, sengupta2018functional, yu2018hardware} as an adversary cannot precisely specify the trigger conditions when a design is locked. However, we exploit a Trojan to obtain the secret key, which is the primary contribution of this paper. 
    
    \item We present three types of TAAL attacks that extract the secret key differently using hardware Trojans placed at different locations in the netlist. \textit{T1 type TAAL} attack directly leaks the secret key to the primary output of an IC once the Trojan is activated (see Figure~\ref{fig:TAAL-attack-v1_v2}.(a)). On the other hand, \textit{T2 type TAAL} and \textit{T3 type TAAL} attacks rely on the activation and propagation of the secret key to the primary output (see Figure~\ref{fig:TAAL-attack-v1_v2}.(b) and Figure~\ref{fig:TAAL-attack-v3} respectively). Note that an adversary has the freedom of choosing one of these three types of TAAL attacks implemented using combinational, sequential, or analog hardware Trojans. 
    
    \item We present an existing well-known model for designing a combinational hardware Trojan~\cite{zhou2018modeling}, which can be used to launch the TAAL attacks. An adversary has the freedom to choose the type of hardware Trojan, which can be designed so that it evades manufacturing or production tests and remains undetected. These combinational Trojans can be described as \textit{Type-p} Trojans, as they have $p$ trigger inputs. These triggers can come from the primary inputs and/or internal nodes of a locked circuit, which are not affected by the keys. We present that a very large number of Trojans can be created, and an adversary can use only one such a Trojan. It is practically impossible for an SoC designer to detect all these feasible Trojans using logic tests. 
    
    \item We also present a model for sequential Trojan, which is constructed using a combinational one. A state element (a counter) is added to a combinational Trojan to deliver the payload once it is triggered $R$ times consecutively. Note that the trigger inputs for both the Trojans are the same. The combinational Trojan delivers the payload once triggered; on the other hand, a sequential Trojan needs to be triggered $R$ times consecutively. Note that an adversary can choose any existing design of a hardware Trojan proposed so far.      
    
\end{itemize}

The rest of the paper is organized as follows: an introduction to logic locking along with an overview of different locking techniques and attacks is provided in Section~\ref{sec:background}. The proposed attacks based on hardware Trojan to implement the malicious design modification for the extraction of the secret key from any locked circuit are described in Section~\ref{sec:proposed-attack}. We provide an algorithm for designing an \textit{Type-p} combinational Trojan considering the set of manufacturing test patterns in Section~\ref{sec:hardware-Trojans}. The number of valid Trojans, along with other factors such as area overhead and leakage power for several benchmark circuits, are presented in Section \ref{sec:results}. The future directions are provided in Section~\ref{sec:future-directions}. Finally, we conclude our paper in Section \ref{conclusion}.

\section{Background} \label{sec:background}
The challenges for protecting a circuit against hardware security threats have been the driving force for developing different techniques to limit the amount of circuit information that can be recovered by an adversary. Logic locking has emerged as a field of significant interest from the researchers, as it can provide complete protection against IC overproduction and IP piracy. Different researchers have proposed to use logic locking to prevent hardware Trojans as well~\cite{yasin2015improving, yasin2015transforming, liu2014embedded, yu2017exploiting, sengupta2018functional, yu2018hardware}. 

\begin{figure}[t]
    \centering
    \includegraphics[width=0.8\linewidth]{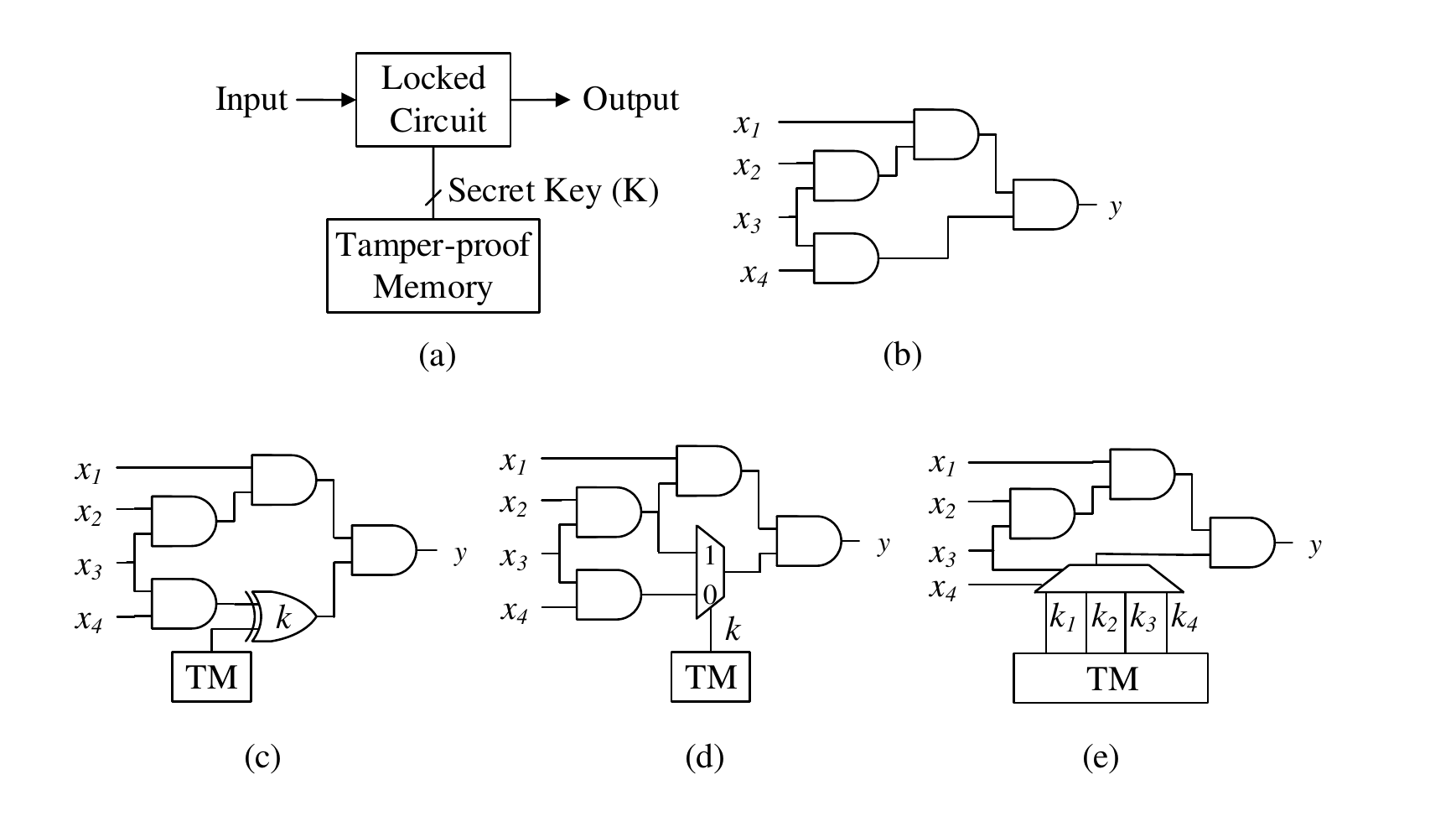}
    \caption{Different logic locking techniques. (a) A locked circuit, where the secret key ($K$) is programmed in a tamper-proof memory ($TM$). (b) Original circuit. (c) XOR-based locking. (d) MUX-based locking. (e) LUT-based locking.} \label{fig:logic_locking} \vspace{-10px}
\end{figure}

The objective of logic locking is to obfuscate the inner details of the circuit and make it infeasible for an adversary to reconstruct the original netlist. Logic Locking hides the circuit's functionality by inserting additional logic gates into the original design, which we termed as \textit{key gates}. In addition to the original inputs, the locked circuit needs secret key inputs to key gates from on-chip tamper-proof memory (see Figure~\ref{fig:logic_locking}$.(a)$ for details). The correct functionality of the design is obtained when the key inputs receive the proper secret key value. Applying an invalid key to the key gates would result in incorrect functionality of the locked design. Note that for a securely locked circuit, the design details cannot be recovered using reverse engineering. 

Different logic locking methods were devised over the years and can be categorized into three different categories. First, XOR-based logic locking, shown in Figure~\ref{fig:logic_locking}.(c), has received much attention due to its simplicity. In this technique, a set of XOR or XNOR gates are inserted as key gates~\cite{roy2008epic, rajendran2012security, guin2016fortis, xie2016mitigating, yasin2016sarlock,  GuinVTS2017, yasin2017lock, GuinTVLSI2018, yasin2017provably}. The secret key is stored in tamper-proof memory (TM), and connections are made from TM to the key gates. Second, in the MUX-based logic locking technique~\cite{plaza2015solving, lee2015improving}, multiplexers (MUX) are inserted so that one of its input is correct, which is the actual net of the circuit. The other input of the MUX is incorrect, which is a dummy net randomly selected from the netlist. This technique is shown in Figure~\ref{fig:logic_locking}.(d). The select signal of the MUX is associated with the key bit from the tamper-proof memory. The correct signal goes through the MUX upon applying valid key value; otherwise, the incorrect signal propagates in the netlist. Third, in LUT-based logic locking, \cite{baumgarten2010preventing, khaleghi2015ic, liu2014embedded}, shown in Figure~\ref{fig:logic_locking}.(e), a look-up table with several key inputs is used to lock the netlist. The LUTs replace a combinational logic in the design, making it difficult to predict the output as it depends on several different key values.

The research community has proposed several attacks to exploit the security vulnerability on a logic locked circuit. Subramanyan \el \cite{subramanyan2015evaluating} first showed that a locked circuit could be broken using Boolean Satisfiability (SAT) analysis. The SAT attack algorithm, attributed as an oracle-guided attack, requires a locked netlist, which can be recovered using reverse engineering and functional chip with a valid key stored/programmed in its tamper-proof memory. In this attack, an adversary can query an activated chip and observe the response. Note that the SAT attack requires access to the internal nodes of the circuit through the scan chains, which is common in today's netlist for implementing Design-for-Testability (DFT)~\cite{bushnell2004essentials}. The SAT attack works iteratively to eliminate incorrect key values from the key space using distinguishing input patters (DIPs). A DIP is defined as an input pattern for which two sets of hypothesis keys produce complementary results. By comparing these with the output of an unlocked chip, one set of hypothesis keys is discarded. The SAT attack works efficiently as it discards multiple hypothesis keys in one iteration. 

Thereafter, researchers have focused on improving and developing locking techniques to be resilient against the SAT attack. Subsequent work in this direction involved Anti-SAT~\cite{xie2016mitigating,xie2019anti}, SARLock~\cite{yasin2016sarlock}, TTLock~\cite{yasin2017lock}, SFLL~\cite{yasin2017provably}, design-for-security (DFS) architecture~\cite{guin2016fortis, GuinVTS2017,GuinTVLSI2018}. These proposed techniques use one-point functions. SARlock inverts the output of the circuit for one input pattern corresponding to one incorrect key. This input pattern differs for different incorrect keys. Anti-SAT involves two complementary external logic circuits which are supplied with the same key values and same inputs. The output of these two circuits converge into a AND gate whose output is always 0 for the correct key applied; else, it may be 1 that leads to corrupted internal node value in the original netlist to produce incorrect outputs. Anti-SAT, initially, was proven vulnerable to the signal probability skew (SPS)~\cite{yasin2016sarlock} attack and removal/bypass attack~\cite{xu2017novel}. SARLock was broken by Double DIP attack~\cite{shen2017double}, Approximate SAT attack~\cite{shamsi2017appsat} and removal/bypass attack~\cite{xu2017novel}.

Due to limitations of SARlock, Yasin \el developed an improved version of this design and referred to as TTLock~\cite{yasin2017lock}, where the original design itself is modified to produce corrupted/inverted results upon applying an incorrect key for a single input pattern.  Stripped functionality based logic locking (SFLL)~\cite{yasin2017provably} was proposed to provide more flexibility in the number of protected input patterns. The design is no longer the same as the original design due to stripped parts of the functionality resulting in erroneous output. A separate restore unit is responsible for removing this error in the output, supplying correct key values to it. However, Subramanyan \el has shown that SFLL can be defeated through FALL attack~\cite{sirone2018functional}. The attack is built on three primary steps, namely, structural analysis, functional analysis, and key confirmation. The structural analysis is performed to identify the gates that are the output of the cube stripping function in SFLL. After identification of these candidate gates, the functional analysis targets the property of cube stripping functions, which results in a set of potential key values. Finally, the key confirmation algorithm identifies the correct key from the set of potential key values.  

As the SAT-attack is based on the availability of accessing the internal states of a circuit through the scan chains, Guin \el proposed placing multiple flip-flops capturing signals controlled by different key bits at the same level of the parallel scan chains, which were used in the current test compression methodologies~\cite{guin2016fortis}. However, a vulnerability existed in this design, when an adversary performs multi-cycle tests, such as delay tests (transition delay faults and path delay faults)~\cite{bushnell2004essentials}. This leads to the necessity for developing a new design-for-security (DFS) architecture to prevent leaking of the key during any manufacturing tests~\cite{GuinVTS2017, GuinTVLSI2018}. This design prevents scanning out the internal states after a chip is being activated, and the keys are programmed/stored in the circuit. 

Apart from SAT-based attacks, probing attacks~\cite{wang2017probing, rahman2019key} have also shown serious threats to the security of logic locking, where an attacker makes contact with the probes at signal wires in order to extract sensitive information, mainly, the secret key. With the help of a focused ion beam (FIB), a powerful circuit editing tool that can mill and deposit material with nanoscale precision, an attacker can circumvent protection mechanisms and reach wires carrying sensitive information. However, the countermeasures reflect the complexity of shield-structure and nanopyramid structures as the defense, making it difficult to perform these attacks~\cite{wang2019probing,shen2018nanopyramid}. Recently, Zhang \el proposed an oracle less attack to extract the key from locked circuits~\cite{zhang2019tga}. The notion of this attack is to compare the locked and unlocked instances of repeated Boolean functions in the netlist to predict the key. A solution was proposed to countermeasure the attack as well.    

\section{Proposed TAAL Attack for Extracting Secret Keys} \label{sec:proposed-attack}

The general hardware security strategy adopted for designing and manufacturing a circuit involves a logic locking, where a chip is unlocked by storing a secret key in the tamper-proof memory. As this secret key is the same for all the chips manufactured with the same design, finding this key from one chip undermines the security resulted from logic locking. We show that an adversary can easily extract the key for a chip using our proposed TAAL attacks, built on tampering through malicious modification by inserting a hardware Trojan to a locked circuit. In this section, we will present three types of TAAL attacks. 

\subsection{Adversarial Model} \label{subsec:adversarial-model}
The adversarial model is given to defining the capabilities and intentions of an attacker clearly. In this model, the attacker (adversary) is assumed to be an untrusted foundry and possesses the following:
\begin{itemize}
    \item The attacker has access to the locked netlist of a circuit. An untrusted foundry has access to all the layout information, which can be extracted from the GDSII or OASIS file.  The netlist can be reconstructed from the layout using reverse engineering with advanced technological tools~\cite{torrance2009state}.
    
    \item The attacker can determine the location of the tamper-proof memory. It can also find the location of key gates in a netlist, as it can easily trace the route of the other input of the key gate to the tamper-proof memory.
    
    \item The attacker can tamper a netlist for its malicious intentions through inserting additional circuitry, commonly known as hardware Trojans, about which the SoC designer is unaware. 
    
    \item The attacker has access to all the manufacturing test (\eg stuck-at-fault, delay fault) patterns. Commonly, the production tests are performed at the foundry.   
\end{itemize}

\subsection{T1 Type TAAL Attack} \label{subsec:TAAL-attack-v1} 
The \textit{T1 type TAAL} attack is the most straightforward attack among the other two types, which will be introduced in the successive sections. This attack is beneficial for the attacker who does not intend to gain knowledge regarding the security measures implemented for the circuit. The hardware Trojan assists in extracting out the secret key directly from the tamper-proof memory.    

\begin{figure} [t]
    \centering
    \includegraphics[width=\linewidth]{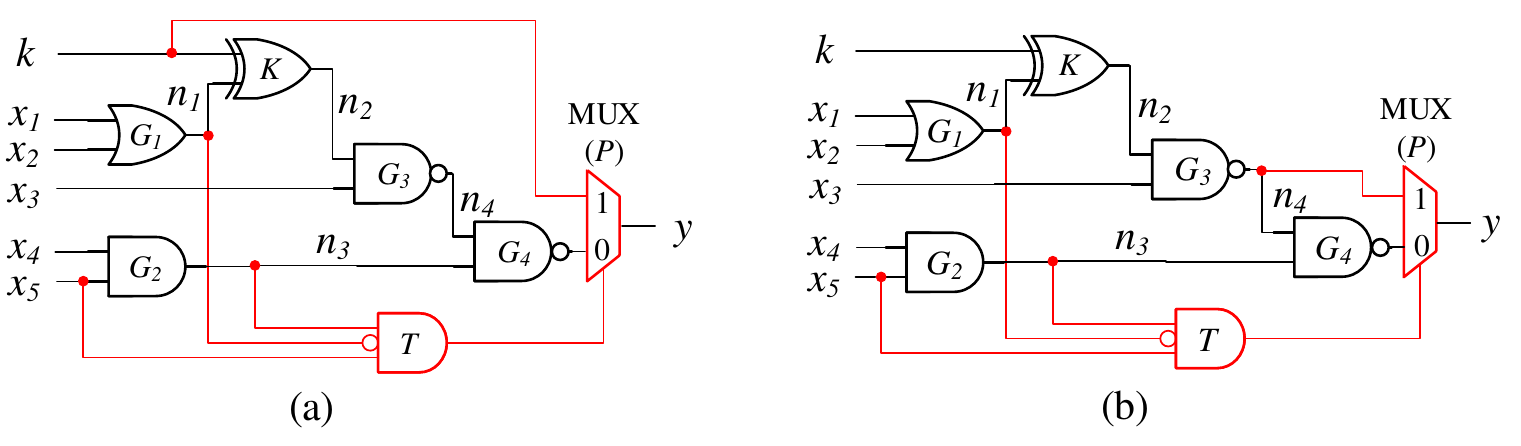}
    \caption{(a) \textit{T1 type TAAL attack}, where a \textit{Type-3} combinational Trojan is inserted for key extraction directly from the connection between key gate and tamper-proof memory, (b) \textit{T2 type TAAL attack}, where a \textit{Type-3} combinational Trojan is inserted for the secret key extraction.}
    \label{fig:TAAL-attack-v1_v2} \vspace{-10px}
\end{figure}

Figure~\ref{fig:TAAL-attack-v1_v2}.(a) shows the proposed modification to launch \textit{T1 type TAAL} attack. A \textit{Type-3} combinational Trojan (see details in Section~\ref{subsec:combinational-hardware-Trojan}) is designed and inserted in the netlist. A combination of $3$-input AND gate with an inverter at one of its input served as the trigger and denoted as $T$, whereas the $2$-input multiplexer delivers the payload to the primary output. One input of the multiplexer is the actual output of the locked netlist. The other input is connected to the line formed between the key gate ($K$) and the tamper-proof memory as a part of logic locking. Under normal operation for any activated chip, the multiplexer propagates the correct circuit functionality at the output. Once the Trojan gets activated with $[x_1~x_2~x_3~x_4~x_5]=[0~0~X~1~1]$, the output of AND gate becomes 1, which leads to the extraction of the secret key through the multiplexer at the output. Note that the required number of multiplexers to extract the complete secret key is dependent on the key size.

The \textit{T1 type TAAL} attack is very effective as it does not require any knowledge of the circuit netlist. This attack can also be applied to any logic locking techniques without knowing its implementation details as it directly leaks the key to the primary output. As there are no security measures undertaken in logic locking to protect the connection between the key gates and the tamper-proof memory, any locked circuit can be vulnerable for this attack. Note that an adversary can select any hardware Trojans (combinational or sequential Trojans) of its choice, and one can find the implementation details in Section~\ref{sec:hardware-Trojans}. %However, an SoC designer can perform reverse-engineering to find out any modification in the netlist by tracing the lines originated from the tamper-proof memory. \todo{Not sure to mention RE here}. Don't delete this comment.

\subsection{T2 Type TAAL Attack} \label{subsec:TAAL-attack-v2}
Instead of extracting the key directly to the primary output, an adversary can propagate it to the output. In \textit{T1 type TAAL Attack}, once the Trojan is activated, the raw key values are transferred to primary output, which can raise a suspicion of a design being tampered. \textit{T2 type TAAL Attack} primarily addresses this shortcoming of \textit{T1 type attack}, by incorporating logic values in the key, which can easily be separated. 

The attack involves tampering a netlist with a \textit{Type-3} combinational Trojan. Similar to \textit{T1 type}, the trigger is constructed using a $3$-input AND gate along with an inverter placed before one of the AND gate inputs (see Figure~\ref{fig:TAAL-attack-v1_v2}.(b)) and the payload is delivered to the primary output of the circuit using a $2$-input MUX. An adversary can choose a net, whose logic value is impacted by the key gate for the input of the MUX. In Figure~\ref{fig:TAAL-attack-v1_v2}.(b), net $n_4$ is selected as the MUX input. One can also select $n_2$, however, $n_3$ cannot be selected. The key gate is considered as XOR gate having inputs as secret key ($k$) and $G_1$ gate output. In order to propagate the secret key at the output of the key gate ($K$), $G_1$ output needs to be specified (either 0 or 1). If $n_1=0$, the output of the key gate will be $k$, otherwise it will be $\overline{k}$. For this example, the trigger requires $n_1=0$; else, the Trojan will not be activated. This condition forces $[x_1~x_2]=[0~0]$. Since net $n_4$ is selected for key extraction, input $x_3$ also plays a significant role in key propagation as the complementary value at net $n_2$ can propagate to $n_4$ only when $x_3$=$1$. To launch this attack, an adversary needs to perform the circuit analysis to sensitize the key. This attack requires an adversary to monitor the input pattern to extract the correct key. An adversary extracts $\overline{k}$ when $[x_1~x_2~x_3~x_4~x_5]=[0~0~1~1~1]$, in the example provided in Figure~\ref{fig:TAAL-attack-v1_v2}.(b). This attack shows the flexibility to identify individual key gate and target secret key through key propagation from consecutive node selection. The dependency of this attack on primary inputs increases the efficiency of \textit{T2 type} attack where only the adversary knows the logical values of these inputs.

\subsection{T3 Type TAAL Attack} \label{subsec:TAAL-attack-v3}
A locked netlist typically consists of a large number (e.g., 128) of key gates, the effect of one key may affect the propagation of another key to the primary output. A secure logic locking technique can also insert keys in such a way that an adversary cannot propagate the key information to output using manufacturing tests \cite{rajendran2012security}. In such scenarios, \textit{T2 type TAAL attack} may be ineffective in extracting the key. \textit{T3 type TAAL attack} is proposed  to addresses this limitation encountered for \textit{T2 type attack}.

\begin{figure}[ht]
     \centering
     \includegraphics[width=\linewidth]{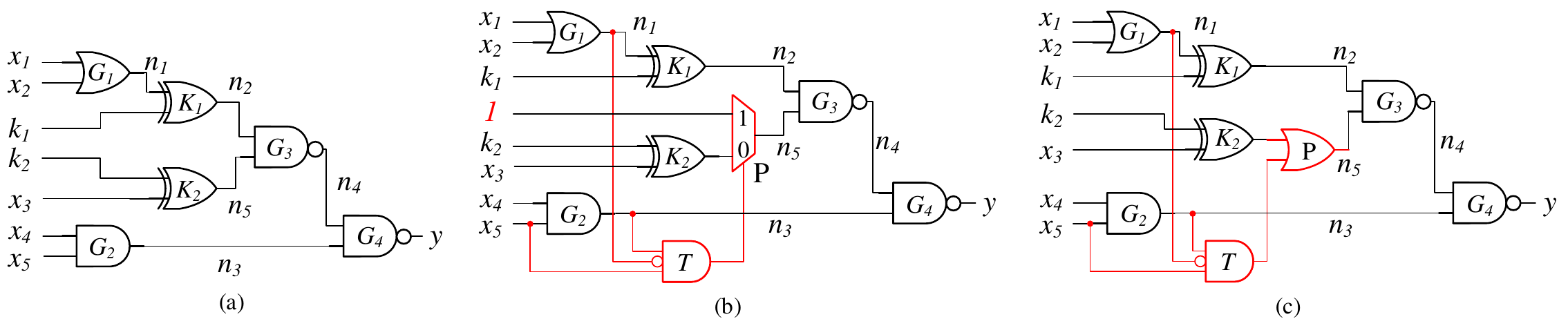}
     \caption{T3 Type TAAL attacks. (a)~Original netlist with $k_2$ inserted to prevent the propagation of $k_1$, (b)~\textit{T3 type TAAL attack} with a \textit{Type-3}  combinational Trojan with payload as multiplexer (MUX) and (c) \textit{T3 type TAAL attack} with a \textit{Type-3} combinational Trojan with payload as OR gate. 
     } 
     \label{fig:TAAL-attack-v3} %\vspace{-10px}
 \end{figure}

Figure~\ref{fig:TAAL-attack-v3}.(a) shows the locked netlist, where the propagation of the key ($k_1$) is prevented by inserting another key ($k_2$). The output of $G_3$ cannot be uniquely determined unless an adversary knows either $k_1$ or $k_2$ and \textit{T2 type attack} will fail to determine either $k_1$ or $k_2$. It is thus necessary to help propagate one key and then determine the other. Figure~\ref{fig:TAAL-attack-v3}.(b) shows our proposed \textit{T3 type TAAL attack}, where net $n_5$ is selected to deliver the payload.

Figure~\ref{fig:TAAL-attack-v3}.(b) shows the implantation of \textit{T3 type TAAL attack} using a \textit{Type-3} combinational Trojan. The trigger part for the Trojan can be designed as $3$-input AND gate with inverter and payload is delivered through $2$-input MUX as before. The key ($k_1$) propagation requires setting the node $n_5$ to 1 so that the signal value at node $n_2$ can propagate through $G_3$. The other input of the MUX is directly connected to $V_{DD}$, which is equivalent to logic 1. 

The other input of the key gate ($K_1$) needs to be specified to propagate the key $k_1$ at node $n_2$. As one of the inputs to trigger requires $n_1=0$ to be satisfied, which results $[x_1~x_2]=[0~0]$, and the output of key gate ($K_1$) will be $k_1$. When the trigger condition is met, the output of the AND gate becomes 1, and logic 1 is delivered at the input of $G_3$ through the MUX. At this point, the Trojan activation nullifies the effect of key value $k_2$ at the output of $G3$. The signal at $n_4$ will be $\overline{k_1}$. Finally, setting node $n_3$ at logic 1 will expose the key at the output, $y$. As a result, input pattern $[x_1~x_2~x_3~x_4~x_5]=[0~0~X~1~1]$ will expose the key $k_1$ at the output. Once the value of $k_1$ is known, an adversary can perform the signal propagation analysis to find $k_2$. Similarly, the payload MUX can be replaced with an OR gate as shown in Figure~\ref{fig:TAAL-attack-v3}.(c), the attack works in the same manner where triggering of Trojan would force the output of the payload OR gate to logic 1 irrespective of its other input which will assist in propagating the key value $k_1$ at the primary output depending on the primary input values as discussed above.

The attacks are explained using a combinational Trojan for the simplicity of understanding. All the attacks proposed can also be implemented using any \textit{Type-p} sequential Trojan, which delivers at its targeted payload once the Trojan is activated $R$ times. The design details for a sequential Trojan is discussed in Section~\ref{subsec:sequential-hardware-Trojan}.

Note that the \textit{T3 Type} TAAL attack cannot be applied to DFS structure~\cite{GuinVTS2017, GuinTVLSI2018} since DFS architecture prevents leaking of key through the secure cell. However, the adversary can implement either \textit{T1 Type} and \textit{T2 Type} TAAL attacks to bypass the secure cell and be extracted out through scan-chain.

\section{Design of Hardware Trojans for TAAL Attacks} \label{sec:hardware-Trojans}
A hardware Trojan can be described as intentional modifications in the original netlist of a design for malicious purposes~\cite{adee2008hunt, tehranipoor2010survey, karri2010trustworthy, tehranipoor2011trustworthy, bhunia2014hardware, sturton2011defeating}. A Trojan can be inserted into a circuit during its design or manufacturing stages. In this paper, we only consider a Trojan, inserted by an untrusted foundry, which is relevant to logic locking (see the adversarial model in Section~\ref{subsec:adversarial-model}). As logic locking was proposed to address the threat from an untrusted foundry, it can practically thwart the locking mechanism by obtaining the secret key through inserting a hardware Trojan in the design.

A complete hardware Trojan classification can be found in~\cite{bhunia2018hardware}. In this paper, we only consider combinational and sequential hardware Trojans to demonstrate the attack. A combinational hardware Trojan generally comprises of a trigger and a payload, the detailed modeling can be found in~\cite{zhou2018modeling}. On the other hand, sequential Trojans have a state element along with the trigger and payload~\cite{bhunia2014hardware,wang2011sequential}. Any Trojans can be activated through trigger inputs, which can be taken from the primary inputs and/or internal nodes of a circuit so that manufacturing test patterns cannot trigger a Trojan and remain undetected. The trigger can be implemented as an AND gate. When a Trojan is activated, the output of this AND gate becomes 1 and it delivers the payload (selection input of the multiplexer shown in Figures~\ref{fig:TAAL-attack-v1_v2}-\ref{fig:TAAL-attack-v3}) to the circuit to leak the secret key. The trigger can also be any logic function that provides 1 when activated. Note that a combinational Trojan manifests its effects upon the availability of the trigger inputs and effects the original netlist at the payload, on the other hand, sequential Trojan shows its effect after the occurrence of a sequence or a period of time upon triggered.

\subsection{Design for a Combinational Hardware Trojan} \label{subsec:combinational-hardware-Trojan}

The primary purpose of a hardware Trojan, once activated, is to modify the original functionality of a circuit to leak the secret key, which is unknown to the SoC designer. It is absolutely necessary that the Trojan must not get activated during scan-based structural or functional tests. In other words, the circuit should not come across any condition during tests that activates the trigger, which can lead to its detection. In this paper, we present a step-by-step process for designing a combinational hardware Trojan, initially presented in~\cite{zhou2018modeling}, for a locked netlist. 

\begin{figure}[t]
     \centering
     \includegraphics[width=\linewidth]{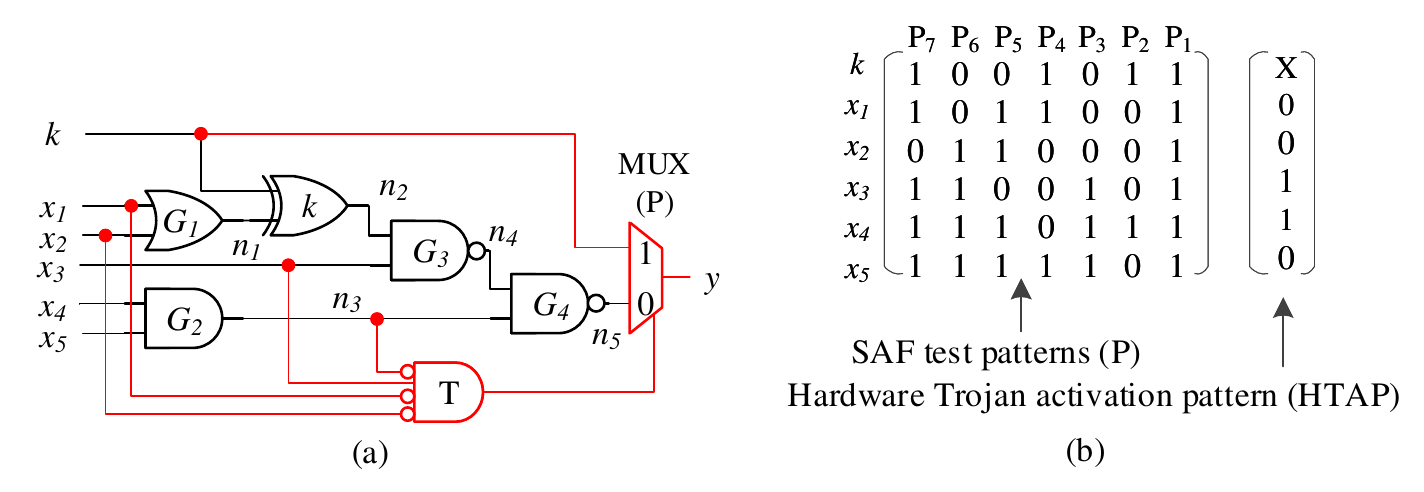}
     \caption{ Design for a combinational hardware Trojan that evades manufacturing tests. (a)~A combinational circuit with a \textit{Type-4} Trojan. (b) Stuck-at fault (SAF) test patterns for manufacturing tests. The hardware Trojan activation pattern (\textit{HTAP}) is $[x_1~x_2~x_3~x_4~x_5]=[0~0~1~1~0]$ while treating the key input as unknown (X).}
     \label{fig:Trojan_trigger}
 \end{figure}

A hardware Trojan can be described based on its trigger inputs, and can be defined as \textit{Type-p} Trojan when it has \textit{p} trigger inputs. The trigger inputs can be selected from primary inputs and/or internal nodes, which are not affected by the key gates. If such a node is selected as a trigger input, an adversary cannot activate a Trojan as it does not know this secret key, and thus the internal signal value for an activation pattern. We call this pattern as hardware Trojan activation pattern (HTAP). The payload of the Trojan (a MUX) can be delivered to a location described in Figures~\ref{fig:Trojan_trigger} for launching our proposed TAAL attack. 

Let us determine the number of Trojans, one can insert in a design to extract the secret key. This is basically a selection problem, where an adversary selects $p$ nodes as the trigger inputs from $N$ nodes of a circuit so that the Trojan is not activated during the manufacturing/production tests. The value of $N$ can be determined based on the following equation:

\vspace{-10px}
\begin{equation} \label{eq:no-of-trigger-nets}
 N = PI + G + F -M  
\end{equation} \vspace{-10px}

\begin{tabular}{cccl}
where,  &    &   & \\
        & PI & : & Number of primary inputs \\
        & G  & : & Number of gates \\
        & F  & : & Number of fanout branches \\
        & M  & : & Number of lines impacted by the\\
        &    &   & key gates
\end{tabular} \vspace{5px}

An upper bound of all possible \textit{Type-p} Trojans ($AT_p$) can be given by:

\begin{equation} \label{eq:all-Trojans}
 AT_p = {N \choose p} \times 2^p 
\end{equation}

The right-hand side of Equation \ref{eq:all-Trojans} constitutes of two products. The first one represents all possible combinations to select \textit{p} lines from \textit{N}. The second one denotes the trigger combinations, as one line can be applied directly or inverted to the trigger input. Note that the actual number of Trojans (denoted as $VT_p$) can be less than $AT_p$ as few of them can be detected by the manufacturing test patterns (e.g., stuck-at fault patterns), and few may not be triggered from the primary inputs. However, for a reasonable size circuit, $AT_p$ and $VT_p$ are comparable.  

Figure \ref{fig:Trojan_trigger} shows an example of \textit{TAAL: T1 type} using a \textit{Type-4} Trojan inserted in the netlist. The circuit has five primary inputs ($PI$). The SoC designer can generate test patterns considering the key as input (the pattern generation is described in detail in~\cite{GuinVTS2017, GuinTVLSI2018}. To detect all the stuck-at faults (SAFs), seven test patterns (\eg, $P=\{P_1$, $P_2$, \ldots, $P_7\}$) are required and they are generated using Synopsys TetraMax~\cite{SynopsysTetraMAX} ATPG tool. To avoid a Trojan being activated by these manufacturing test patterns, the Trojan's trigger must remain quiet for all these input patterns. A hardware Trojan activation pattern is selected, where $HTAP=(X~0~0~1~1~0)^T \notin P$. As the logic values of nodes $n_2$, $n_4$, and $n_5$ are impacted by the key, $k$, these nodes are excluded in designing the Trojan. If one of these nodes is selected, an adversary may not activate the trigger as it does not know the key value. The upper bound of all possible \textit{Type-4} Trojans ($AT_4$) can be given by:

\begin{equation}
    AT_4 = {7 \choose 4} \times 2^4 = 560
\end{equation}
where, $N=5+5-3=7$.

%\todoZiqi{how many of them will be detected by P?   Uploaded.}
%\todo{update the number for valid Trojans- Ayush}

Out of these 560 possible Trojans, 188 will be detected by the test patterns $P$. The remaining 372 will be treated as valid Trojans, and an adversary can select one of them. In Figure \ref{fig:Trojan_trigger}, $\overline{x_1}$, $x_2$, $\overline{x_3}$ and $\overline{n_3}$ are selected as the trigger. Similarly, one can also design other types (\textit{Type-1} through \textit{Type-6}) of Trojans to launch TAAL attacks. Note that the Trojan needs to be quiet during normal operations so that no functional errors are observed at the primary output. An adversary can select rare nodes, whose value do not become identical with trigger pattern very often under continuous/normal operation of the IC, for the trigger inputs while designing a Trojan. One can perform controllability, and observability analysis \cite{bushnell2004essentials} to find such rare nodes, and then select them as the trigger inputs. However, we do not include such analysis, as including rare nodes in the trigger for a sequential hardware Trojan is not a hard requirement. In this paper, a sequential Trojan is modeled using a combinational Trojan that needs to be triggered $R$ times consecutively.

\begin{algorithm}
\SetAlgoLined
\SetKwInOut{Input}{input}\SetKwInOut{Output}{output}
\Input{~Locked Netlist ($C$), test pattern set ($P$), \textit{Type-p Trojan}}
\Output{~Hardware Trojan activation pattern (\textit{HTAP}), Trigger inputs ($T$)} 
\vspace{3px}
%\hline
\vspace{3px}
Read the locked netlist ($C$)\;
Read production test patterns ($P$)\;

Perform logic simulation using $P$ to form a matrix ($A$) of all the internal node values\;

Select a hardware Trojan activation pattern ($HTAP$), where $HTAP \not\in \ P$ \;

Perform logic simulation using $HTAP$ and form a matrix ($H$) of all the internal node values\;

Select $p$ random nodes that are not affected by the key gates of \textit{C} for the trigger inputs\;

Construct a new matrix $A_p$ that corresponds to the trigger locations for all test patterns\;

Construct a new vector $H_p$ that corresponds to the trigger locations for $HTAP$ \;

  \eIf{$H_p \notin A_p$}{
   Choose selected $p$ nodes as trigger, $T$\; 
   }{
   Discard selected $p$ nodes, as it would activate the Trojan during tests\;
   Go to Step 6\;
  }
  
Report $HTAP$ and $T$\;
\caption{Design of Type-$p$ Trojan} \label{alg:design}
\end{algorithm}

An automated process is developed to design a combinational hardware Trojan. Algorithm \ref{alg:design} provides steps to be followed for designing a \textit{Type-p} Trojan that eludes activation during the manufacturing test. The inputs of the algorithm are locked netlist($C$), $M$ production/manufacturing test patterns ($P=\{P_1,~P_2,~\ldots,~P_M\}$) and the Trojan type ($p$). The algorithm results in the hardware Trojan activation pattern (HTAP) and the trigger inputs. Initially, it reads the Trojan-free locked netlist ($C$) and the set of manufacturing test patterns ($P$) (Lines 1-2). Logic simulation is performed using these patterns, and store the internal node values in a matrix, $A$ (Line 3). It is unnecessary to store the nodes that are impacted by the key gates, as their values will be unknown (Xs) during the simulation. Note that $A$ is a $N \times M$ matrix. A hardware Trojan activation pattern of an adversary's choice is selected (Line 4). Similarly, matrix $H$ is formed due to logic simulation using HTAP (Line 5). Here, $H$ is a $N\times1$ vector. To select the trigger inputs, one can select $p$ random nodes that are not affected by the key gates of the locked circuit (Line 6). To perform the search whether the trigger values are presented in the production set, the matrix $A_p$ and vector $H_p$ are constructed (Lines 7-8). If $H_p$ is not in $A_p$, the trigger ($T$) is selected (Line 10), otherwise, drop the selected $p$ locations as the Trojan will be activated during the production tests (Lines 12-13) and new $p$ locations are selected (Line 6). Finally, the algorithm reports $HTAP$ and $T$ (Line 15). 

\subsection{Design for a Sequential Hardware Trojan} \label{subsec:sequential-hardware-Trojan}

A sequential Trojan modifies the functionality of a circuit until a specified time has elapsed after the trigger condition is satisfied. However, in this paper, we designed a sequential Trojan that needs to be triggered $R$ times to deliver the payload. We designed a sequential Trojan in this way so that it can be modeled using a combinational Trojan, described in detail in the previous section.  

\begin{figure}[t]
    \centering
    \includegraphics[width=\linewidth]{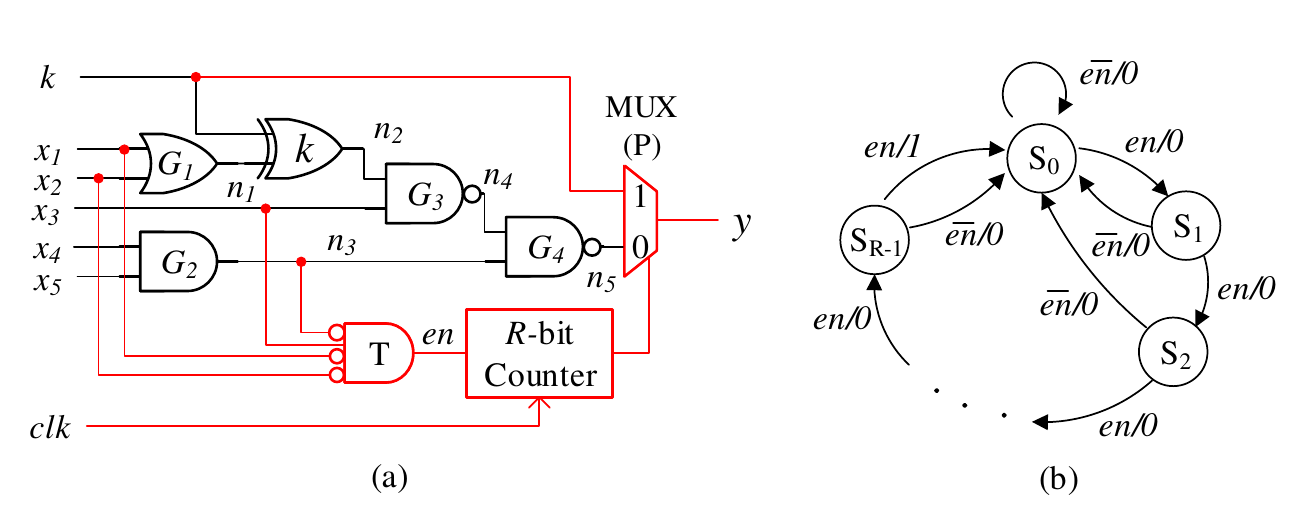}
    \caption{(a) The netlist of a sequential Trojan with a \textit{R-bit} counter, (b) The finite state machine (FSM) of the counter used in a sequential Trojan.}
    \label{fig:Sequnetial Trojan} %\vspace{-10px}
\end{figure}

A sequential Trojan also consists of a trigger and payload similar to a combinational Trojan. Additionally, the trigger part contains state elements that ascertain the payload in future time. In our sequential Trojan design, a \textit{R-bit} counter is implemented as the state elements. This counter is enabled ($en$) once the trigger condition is fulfilled, i.e., the output of the \textit{p}-input AND gate becomes 1. The counter increments by one unit, every-time the Trojan is triggered using the Trojan activation pattern (HTAP). The Trojan delivers at the payload (MUX) only after reaching the maximum count value ($R$). The circuit tampered with Sequential Trojan will show the intended malfunction, key extraction in TAAL attacks, only upon applying the activation pattern successively \textit{R}-times to the circuit.

Figure~\ref{fig:Sequnetial Trojan}.(a) shows the \textit{TAAL attack} using a sequential Trojan. The trigger consists of a $p$-input AND gate and a $R$-bit counter. The finite-state machine (FSM) of the counter is shown in Figure~\ref{fig:Sequnetial Trojan}.(b). The FSM goes to the next states when $en=1$; otherwise, it returns to the initial state, $S_0$. The counter produces an output of 1, once $en$ is hold to 1 consecutively $R$ clock cycles, as it takes $(R-1)$ cycles to reach $S_{R-1}$. Note that this sequential Trojan can be modeled as \textit{R} number of combinational Trojans. An adversary can also design a different sequential Trojan, already in the literature, to launch the \textit{TAAL attack}. The sequential Trojan increases the complexity compared to a combinational Trojan as it manifests its effect to the payload only after the sequence of repeated application of trigger inputs. Only the adversary has the knowledge regarding the maximum counter value making it very difficult for detection. 

%{\color{red}
Note that the Trojan flip-flops (\textit{FFs}) cannot be a part of the scan chain, since a Trojan is implanted after the test pattern generation (if this process is performed at the design house). Any modifications in the scan chain will be detected with a stuck-at fault test. As the Trojan circuitry is extremely small (one counter and AND gate for the Trigger), it is possible to verify the proper operation using a functional test (i.e., applying the Trigger pattern $R$ times and observe the response). There is no need to add Trojan \textit{FFs} in the scan chain and perform scan tests. Now, one can argue that test patterns are generated at the foundry, and Trojan \textit{FFs} are a part of the scan chain. In such a situation, the scan shift will not change the state of the original \textit{FFs} in the circuit. However, it is highly unlikely that an adversary will add these malicious \textit{FFs} in the scan chains, as one can easily find the response of the combinational part of the Trojan and determine tampering.
%}

\subsection{Design for an Analog/RF Trojan} \label{subsec:Analog-RF-TRojan}
Analog Trojans~\cite{yang2016a2,nagarajan2019entt,kison2019security} can also be used to launch different TAAL attacks. In this section, we provide a brief description of different analog Trojans. The payload can be implemented using a multiplexer or an OR gate (see Figures~\ref{fig:TAAL-attack-v1_v2}-\ref{fig:TAAL-attack-v3}) like a combinational or sequential hardware Trojan. The implementation of the trigger, however, be different depending on the Trojan design. The analog or RF-leaking Trojans can detect an extremely rare event with just a handful of transistors added to the circuit. Moreover, analog techniques and characteristics to design the stealthy trigger for Trojans are usually difficult to uncover due to their small footprint~\cite{yang2016a2}. The authors used the switched capacitor to design a trigger circuit, which is activated when the frequency of toggling on a victim wire goes above a certain threshold. When the wire toggles frequently, charge accumulates on the capacitor faster than it leaks away, eventually the voltage of the capacitor rises above the threshold. This deploys the payload to cause intentional malicious activity. A similar notion is utilized to introduce triggers that are activated after some delay or operate on a specific voltage threshold~\cite{nagarajan2019entt}. Kison \el exploited the capacitor coupling in sub-micron process technologies to design Trojans~\cite{kison2019security}. This technique uses rerouting and extending existing layout tracks to increase the capacitive coupling between a victim and an aggressor wire in a way that a low to high transition on the aggressor can adequately affect the victim wire and flip its digital value. Similarly, RF-leaking Trojans leak the information through the Trojan-induced channel without affecting the legitimate signal/channel~\cite{subramani2017ace, chang2015enabling}.
%}

\section{Analysis}\label{sec:results}
The hardware Trojans presented in Section~\ref{sec:hardware-Trojans} pose a unique challenge to the SoC designers for securing their designs. 

\subsection{Complexity Analysis}
In this section, we show that an adversary can implement a Trojan in a very large number of ways, and it is practically infeasible to detect all of them with absolute certainty.  We choose six benchmark circuits from ISCAS'85 benchmark suites~\cite{bryan1985iscas} to show the complexity of Trojan detection even for these small benchmark circuits.  

\begin{table*}[!ht]
\centering 
\caption{Circuit parameters.} \label{tab:circuit-parameters}
\begin{tabular}{|M{2cm}|M{1.55cm}|M{1.55cm}|M{1.55cm}|M{1.55cm}|M{1.55cm}|M{1.55cm}|}
\hline
\bf Benchmarks  & \bf \# Gates & \bf Key Size ($|K|$) & \bf \# Total Lines ($N+M$) & \bf \# Net Lines ($N$) & \bf \# Test Patterns &\bf Fault coverage \\ \hline
C432      & 160   & 30  & 349    & 233  & 58 & 100\%  \\ \hline
C499      & 202   & 30  & 491    & 226  & 78  & 100\%  \\ \hline
C880      & 383   & 30  & 594    & 350  & 86  & 100\%  \\ \hline
%C1355    & 733   & 30  & 731  &  \todoZiqi{???}& 125 &  99.15\% \\ \hline
C1908     & 880   & 30  & 552    & 223  & 83  & 100\%  \\ \hline
%C2670     & 1269  & 124  & 99.89\%  &  &  &   \\ \hline
C3540     & 1669  & 83  & 1826   & 1114 & 173  & 100\%  \\ \hline
%C5315     & 2307  & 119  & 100\%    &  &  &   \\ \hline
C6288     & 2416  & 128 & 5621   & 1335  & 77  & 100\%   \\ \hline
\end{tabular}
%\vspace{-10px}
\end{table*}

Table~\ref{tab:circuit-parameters} shows the design details for different locked benchmark circuits. The number of logic gates and key size for these circuits is shown in Columns 2 and 3. The number of key bits is selected in such a way that the total area overhead does not exceed 5\%. However, for industrial design with millions of gates, the key gates will merely add any overhead. Column 4 represents the total number of nets in these circuits (see Equation~\ref{eq:no-of-trigger-nets}). The number of nets that are not affected by key gates is shown in Column 5. A key gate is selected by analyzing the netlist and forward tracing is performed till the primary output(s) is reached. Simultaneously, removing all the nodes that belongs to these paths from the overall list of $N+M$. Note that this includes any fanout branches as well while performing forward tracing. This step is repeated for all the key gates to get the nodes that are not impacted by the key values. Note that these nets cannot be selected for trigger inputs. The manufacturing test patterns are generated using Synopsys TetraMax Automatic Test Pattern Generation (ATPG) tool~\cite{SynopsysTetraMAX} with targeted 100\% fault coverage (Columns 6-7). For the C432 benchmark, we insert 30 key gates randomly in the netlist with 160 logic gates. There are 349 nets in the netlist, out of which 233 nets can be selected for the Trojan trigger as the remaining nets are affected by the key gates. The TetraMax ATPG tool generates 58 stuck-at fault patterns and reports 100\% fault coverage. An adversary will use these test patterns to design the Trojans such that they are not activated during the manufacturing tests. A similar analysis can be performed for all other benchmark circuits through the details mentioned in respective rows.

\begin{table*}[!ht]
\centering
{\linespread{1} \selectfont

\caption{Number of hardware Trojans for launching TAAL attacks.} \label{tab:complexity}
\begin{tabular}{|M{2.1cm}||M{1.5cm}|M{1.5cm}||M{1.5cm}|M{1.5cm}||M{1.5cm}|M{1.5cm}|}
\hline
\multirow{2}{*}{\textbf{Benchmarks}} & \multicolumn{2}{c||}{\textbf{Type-2 Trojan}} & \multicolumn{2}{c||}{\textbf{Type-3 Trojan}} & \multicolumn{2}{c|}{\textbf{Type-4 Trojan}} \\ \cline{2-7}
 & \bm{$AT_2$} & \bm{$VT_2$} & \bm{$AT_{3}$} & \bm{$VT_{3}$} & $\bm{AT_4}$ & \bm{$VT_4$} \\ \hline %& \bm{$AT_{16}$} & \bm{$VT_{16}$} \\ \hline
C432    & $1.08 \times 10^5$   & $1.04 \times 10^5$    & $1.66 \times 10^7$   & $1.43 \times 10^7$    & $1.91 \times 10^9$    & $1.34 \times 10^9$ \\ \hline
C499    & $1.02 \times 10^5$   & $0.27 \times 10^5$    & $1.52 \times 10^7$   & $2.11 \times 10^6$    & $1.69 \times 10^9$    & $1.21 \times 10^8$  \\ \hline
C880    & $2.44 \times 10^5$   & $2.25 \times 10^5$    & $5.67 \times 10^7$   & $4.80 \times 10^7$    & $9.83 \times 10^9$    & $7.35 \times 10^9$\\ \hline
%C1355   &               & & & & & & &  \\ \hline
C1908   & $0.99 \times 10^5$   & $0.96 \times 10^5$    & $1.46 \times 10^7$   & $1.33 \times 10^7$    & $1.60 \times 10^9$    & $1.27 \times 10^9$\\ \hline
C3540   & $2.48 \times 10^6$   & $2.35 \times 10^6$    & $1.84 \times 10^9$   & $1.57 \times 10^9$ & $1.02 \times 10^{12}$ & $0.74 \times 10^{12}$\\ \hline
C6288   & $3.56 \times 10^6$   & $3.50 \times 10^6$    & $3.17 \times 10^9$   & $3.00 \times 10^9$ & $2.11 \times 10^{12}$ & $1.82 \times 10^{12}$\\ \hline
%& & & & & & & & & & & & \\ \hline
%& & & & & & & & & & & & \\ \hline
%& & & & & & & & & & & & \\ \hline
\end{tabular}
} %\vspace{-10px}
\end{table*}

Table~\ref{tab:complexity} shows the number of combinational hardware Trojans that can be designed to perform \textit{TAAL} attacks (mentioned in Section~\ref{sec:proposed-attack}) for different benchmark circuits. The upper bound (see Equation~\ref{eq:all-Trojans}) for all possible Trojans that can be inserted in the circuit is denoted in Column 2, 4, and 6. Out of all possible Trojans, the valid Trojans that will not be detected during manufacturing tests are shown in Columns 3, 5, and 7. For C432 benchmark circuit, the total number of \textit{Type-2} Trojans is $1.08 \times 10^5$, whereas, the number of valid Trojans is $1.0 \times 10^5$. The number of Trojans increases exponentially with the increase of the Trojan type ($p$). Note that $AT_p$ and $VT_p$ are in the same order, which gives an adversary to select a Trojan of its choice from a large collection. It is worthwhile to mention that an adversary needs to choose a Trojan whose triggers are selected from the rare nodes such that it does not get activated during normal operation. However, it is not necessary to impose this condition for designing a sequential Trojan, as it is highly unlikely that a particular trigger condition will arrive $R$ times consecutively during the normal operation of a chip.   

Note that the result in Table 2 does not show the complexity for sequential hardware Trojans as the ISCAS'85 benchmark circuits are combinational in nature. This table is intended to show the number of possible combinational Trojans, resulted from an effective $N$ number of nets. One can easily extend the results for industrial designs, where an adversary can insert any type of combinational, sequential, or analog Trojans to perform a TAAL attack. Note that we choose small ISCAS'85 benchmark circuits instead of larger benchmarks (e.g., opencores) to demonstrate the difficulty of detecting all the Trojans even for smaller circuits. It is obvious to infer that the difficulty will increase for larger circuits. The number of all possible valid Trojans is in the order of $10^{12}$, which is so large even for a small benchmark circuit, \textit{C6288}, with 2416 gates, when considering only four trigger inputs. If we show that detecting all possible valid Trojans for a very small benchmark is a complex problem, it will be sufficient to maintain that complexity for large circuits.

%{\color{red}
\subsection{Overhead Analysis}%}
\noindent The area for a hardware Trojan can be varied based on the trigger inputs. A \textit{Type-p} combinational Trojan consists of an AND gate with \textit{p}-trigger inputs and a multiplexer or an OR gate as payload. For a sequential Trojan, it is necessary to add a $R$-bit counter along with a $p$-input AND gate to implement the trigger. %{\color{red}
This subsection presents the area and power overhead for different ITC'99 benchmark circuits~\cite{davidson1999itc} locked with a 128-bit key. The simulation is performed by using Synopsys design compiler~\cite{SynopsysDC} with 32nm technology~\cite{Synopsyslibrary}. A single sequential hardware Trojan of \textit{Type-2} to Type-4 is added to each benchmark circuits for launching the \textit{T3 Type} TAAL attack. The output of a single trigger is routed to 128 payload locations (OR gates). This is the worst-case area overhead, as we often need less than 128 OR gates to expose all the 128 key bits. For example, shown in Figure~3.(c), we need only one OR gate to determine two key bits (e.g., $K_1$ and $K_2$). For a given benchmark circuit, $A_O$ represents the original circuit area, whereas $A_T$ represents the area of its Trojan inserted version. The area overhead ($AO$) is computed using the following formula:

 \begin{equation} \label{eq:overhead-calculation}
          AO=\frac{A_T-A_O}{A_O} \times 100 \%
\end{equation}

\begin{table*}[!ht]
\centering
%\color{red}
\caption{Area overhead for ITC'99 benchmark circuits.}  \label{tab:counter_OH} \small
\begin{tabular}{|M{1.7cm}|M{1.6cm}|M{0.8cm}|M{1.2cm}|M{1.0cm}|M{1.0cm}|M{1.0cm}|M{1.0cm}|}
\hline
\multirow{2}{*}{\textbf{Benchmarks}} & \multirow{2}{*}{\textbf{\# Gates}} & \multirow{2}{*}{\textbf{\# FFs}} & \multirow{2}{*}{\begin{tabular}[c]{@{}c@{}}\textbf{Trojan} \\ \textbf{Type}\end{tabular}} & \multicolumn{4}{c|}{\textbf{Area Overhead (\%)}} \\ \cline{5-8} 
 &  &  &  & \textbf{R=2} & \textbf{R=4} & \textbf{R=8} & \textbf{R=16} \\ \hline
\multirow{3}{*}{b14} & \multirow{3}{*}{2064} & \multirow{3}{*}{215} & Type-2 &4.03  &4.23  &4.41  &4.56  \\ \cline{4-8} 
 &  &  & Type-3 &4.05  &4.23  &4.41  &4.59  \\ \cline{4-8} 
 &  &  & Type-4 &4.06  &4.27  &4.41  &4.62  \\ \hline
\multirow{3}{*}{b15} & \multirow{3}{*}{2722} & \multirow{3}{*}{418} & Type-2 &3.02  &3.16  &3.30  &3.41  \\ \cline{4-8} 
 &  &  & Type-3 &3.03  &3.17  &3.30  &3.43  \\ \cline{4-8} 
 &  &  & Type-4 &3.04  &3.19  &3.30  &3.45  \\ \hline
\multirow{3}{*}{b20} & \multirow{3}{*}{4190} & \multirow{3}{*}{430} & Type-2 &1.92  &2.01  &2.10  &2.17  \\ \cline{4-8} 
 &  &  & Type-3 &1.93  &2.02  &2.10  &2.19  \\ \cline{4-8} 
 &  &  & Type-4 &1.94  &2.03  &2.10  &2.20  \\ \hline
\multirow{3}{*}{b21} & \multirow{3}{*}{4225} & \multirow{3}{*}{430} & Type-2 &1.90  &1.99  &2.08  &2.15  \\ \cline{4-8} 
 &  &  & Type-3 &1.91  &1.99  &2.08  &2.16  \\ \cline{4-8} 
 &  &  & Type-4 &1.91  &2.01  &2.06  &2.17  \\ \hline
 \multirow{3}{*}{b17} & \multirow{3}{*}{8629} & \multirow{3}{*}{1328} & Type-2 &0.89  &0.93  &0.97  &1.00  \\ \cline{4-8} 
 &  &  & Type-3 &0.90  &0.94  &0.97  &1.01  \\ \cline{4-8} 
 &  &  & Type-4 &0.90  &0.94  &0.97  &1.02  \\ \hline
\multirow{3}{*}{b18} & \multirow{3}{*}{} & \multirow{3}{*}{} & Type-2 &0.24  &0.25  &0.26  &0.27  \\ \cline{4-8} 
 &39845 & 3168 & Type-3 &0.24  &0.25  &0.26  &0.27  \\ \cline{4-8} 
 &  &  & Type-4 &0.24  &0.25  &0.26  &0.27  \\ \hline
\multirow{3}{*}{b19} & \multirow{3}{*}{} & \multirow{3}{*}{} & Type-2 &0.12  &0.12  &0.13  &0.13  \\ \cline{4-8} 
 & 78076 & 6337 & Type-3 &0.12  &0.12  &0.13  &0.13  \\ \cline{4-8} 
 &  &  & Type-4 &0.12  &0.12  &0.13  &0.13  \\ \hline
\end{tabular}
\end{table*} 

Table~\ref{tab:counter_OH} shows the area overhead analysis for different ITC'99 benchmark circuits. The number of logic gates and flip-flops (FFs) for each synthesized benchmark circuits are shown in Columns 2 and 3, respectively. The Type of Trojan is represented in Column 4. It follows the same definition of \textit{Type-p} Trojan, where $p$ represents the number of trigger inputs. Columns 5 to 8 indicate the percentage area overhead (computed using Equation~$4$).  The Trojan consists of a counter and a AND gate as the trigger, and 128 OR gate as the payload. The $R$ represents the maximum count for the counter. For example, $R$  = 8 means that the trigger has a maximum count of 8.

We observe an area overhead of less than $5\%$ (\eg b14 with only 2064 gates and 215 FFs) for a small benchmark circuit. However, it becomes significantly small for larger benchmark circuits. Considering \textit{b19}, the percentage area overhead is 0.13$\%$ for \textit{Type-2} sequential Trojan with $R = 8$. The area overhead remains almost constant even with an increased count (\eg $R=16$) and different Trojan Types for a large benchmark circuit. A similar analysis can be performed for all the benchmark circuits as well. Note that the majority of the overhead results from the payload as we require 128 OR gates to determine 128 key bits. 

\begin{table}[] \small
\centering
%\color{red}
\caption{Power overhead for benchmark circuits.} 
\label{tab:power-overhead}
%\begin{tabular}{|c|c|c|c|c|c|c|c|c|c|}
\begin{tabular}{|M{1.7cm}|M{1.5cm}|M{0.75cm}|M{0.75cm}|M{0.75cm}|M{0.75cm}|M{0.75cm}|M{0.75cm}|M{0.75cm}|M{0.75cm}|}
\hline
\multirow{2}{*}{\textbf{Benchmarks}} & \multirow{2}{*}{\textbf{\begin{tabular}[c]{@{}c@{}}Trojan\\Types\end{tabular}}}& \multicolumn{4}{c|}{\textbf{Dynamic Power Overhead (\%)}} & \multicolumn{4}{c|}{\textbf{ Leakage Power Overhead (\%)}} \\ \cline{3-10} 
 &  & \textit{\textbf{R=2}} & \textit{\textbf{R=4}} & \textit{\textbf{R=8}} & \textit{\textbf{R=16}} & \textit{\textbf{R=2}} & \textit{\textbf{R=4}} & \textit{\textbf{R=8}} & \textit{\textbf{R=16}} \\ \hline

\multirow{3}{*}{b14} & \textit{Type-2} & 9.61 & 8.46 & 8.63 & 8.86 & 6.02 & 6.63 & 6.89 & 7.05 \\ \cline{2-10} 
 & \textit{Type-3} & 9.70 & 8.34 & 8.60 & 8.80 & 5.91 & 6.62 & 6.77 & 7.03 \\ \cline{2-10} 
 & \textit{Type-4} & 9.63 & 8.38 & 8.52 & 8.70 & 5.95 & 6.64 & 6.82 & 7.05 \\ \hline

\multirow{3}{*}{b15} & \textit{Type-2} & 10.22 & 8.99 & 9.18 & 9.42 & 4.78 & 5.27 & 5.48 & 5.60 \\ \cline{2-10} 
 & \textit{Type-3} & 10.31 & 8.87 & 9.14 & 9.36 & 4.69 & 5.26 & 5.37 & 5.58 \\ \cline{2-10} 
 & \textit{Type-4} & 10.24 & 8.90 & 9.05 & 9.24 & 4.73 & 5.28 & 5.42 & 5.60 \\ \hline

\multirow{3}{*}{b20} & \textit{Type-2} & 7.42 & 6.53 & 6.67 & 6.84 & 3.01 & 3.32 & 3.45 & 3.53 \\ \cline{2-10} 
 & \textit{Type-3} & 7.49 & 6.44 & 6.64 & 6.80 & 2.96 & 3.31 & 3.39 & 3.52 \\ \cline{2-10} 
 & \textit{Type-4} & 7.44 & 6.47 & 6.58 & 6.71 & 2.98 & 3.33 & 3.41 & 3.53 \\ \hline

\multirow{3}{*}{b21} & \textit{Type-2} & 7.67 & 6.75 & 6.89 & 7.07 & 2.99 & 3.30 & 3.43 & 3.51 \\ \cline{2-10} 
 & \textit{Type-3} & 7.74 & 6.66 & 6.86 & 7.03 & 2.94 & 3.29 & 3.36 & 3.49 \\ \cline{2-10} 
 & \textit{Type-4} & 7.69 & 6.68 & 6.80 & 6.94 & 2.96 & 3.30 & 3.39 & 3.51 \\ \hline

\multirow{3}{*}{b17} & \textit{Type-2} & 3.36 & 2.96 & 3.02 & 3.10 & 1.52 & 1.67 & 1.74 & 1.78 \\ \cline{2-10} 
 & \textit{Type-3} & 3.39 & 2.92 & 3.01 & 3.08 & 1.49 & 1.67 & 1.71 & 1.77 \\ \cline{2-10} 
 & \textit{Type-4} & 3.37 & 2.93 & 2.98 & 3.04 & 1.50 & 1.68 & 1.72 & 1.78 \\ \hline

\multirow{3}{*}{b18} & \textit{Type-2} & 0.24 & 0.21 & 0.21 & 0.22 & 0.38 & 0.42 & 0.44 & 0.45 \\ \cline{2-10} 
 & \textit{Type-3} & 0.24 & 0.21 & 0.21 & 0.22 & 0.37 & 0.42 & 0.43 & 0.44 \\ \cline{2-10} 
 & \textit{Type-4} & 0.24 & 0.21 & 0.21 & 0.21 & 0.38 & 0.42 & 0.43 & 0.45 \\ \hline
 
\multirow{3}{*}{b19} & \textit{Type-2} & 0.12 & 0.10 & 0.11 & 0.11 & 0.20 & 0.22 & 0.22 & 0.23 \\ \cline{2-10} 
 & \textit{Type-3} & 0.12 & 0.10 & 0.11 & 0.11 & 0.19 & 0.21 & 0.22 & 0.23 \\ \cline{2-10} 
 & \textit{Type-4} & 0.12 & 0.10 & 0.10 & 0.11 & 0.19 & 0.22 & 0.22 & 0.23 \\ \hline
\end{tabular}
\end{table}

Table~\ref{tab:power-overhead} shows the power overhead analysis for the same benchmark circuits. We have computed two overhead for the dynamic power and the leakage power using Equation~\ref{eqn:dynamic-power-overhead} and Equation~\ref{eqn:leakage-power-overhead}, respectively. As the Trojan remains quiet most of the time unless triggered, leakage power overhead is of the Trojan designers' concern so that it evades detection.  

 \begin{equation} \label{eqn:dynamic-power-overhead}
          DPO=\frac{DP_T-DP_O}{DP_O} \times 100 \%
\end{equation}
where, $DP_T$ and $DP_O$ represent the dynamic power of the Trojan inserted and Trojan free circuits, respectively.

 \begin{equation} \label{eqn:leakage-power-overhead}
          SPO=\frac{SP_T-SP_O}{SP_O} \times 100 \%
\end{equation}
where, $SP_T$, $SP_O$ represent the leakage power of the Trojan inserted and Trojan free circuits, respectively.

Columns 1 and 2 of Table~\ref{tab:power-overhead} represent different benchmark circuits and Trojan types. Columns 3-6 show dynamic power overhead with a sequential Trojan with $R=$2, 4, 8 and 16, respectively. On the other hand, Columns 7-10 show leakage power overhead with a sequential Trojan with $R=$2, 4, 8 and 16, respectively. For example, For a small benchmark circuit, we observe a dynamic power overhead is around $10\%$ (\eg b15) and leakage power overhead is around $7\%~$(\eg b14).
However, for larger benchmark circuits, the power overhead becomes very small. The percentage dynamic power and leakage power overhead for b19 are 0.10$\%$ and 0.22$\%$ for Type-2 sequential Trojan with R=4, respectively. Note that for a large circuit, an increased count (\eg R=16) and different Types of a Trojan will not have a significant impact on dynamic power and leakage power overhead.
%}

\vspace{10px}
\section{Future Research Direction for Secure Logic Locking} \label{sec:future-directions}
The security of a logic locking technique can be tied together with the hardware Trojan detection problem. Developing a SAT-registrant logic locking is not sufficient enough to prevent IC overproduction or to protect IPs. It is required to address the detection of Trojans inserted at an untrusted manufacturing site. Researchers have already proposed different techniques to detect and prevent hardware Trojans. The detection methods can be grouped into two different categories, such as, logic testing~\cite{lesperance2015hardware, waksman2013fanci, chakraborty2009mero, banga2011odette, haider2017advancing}, and side-channel analysis~\cite{agrawal2007trojan, banga2008region, banga2009novel, rad2008sensitivity, li2008speed, liu2014hardware}. On the other hand, prevention methods can be categorized as design-for-trust measures~\cite{chakraborty2009security, rajendran2011design, salmani2012novel, xiao2013bisa, ngo2015linear} and split manufacturing~\cite{vaidyanathan2014detecting, rajendran2013split, wang2016cat}.  

Logic testing by applying stimuli to primary inputs (PIs) and observe responses at primary outputs (POs) can be used to detect these Trojans~\cite{chakraborty2009mero, banga2011odette, Ozgur2013, bhunia2014hardware, lesperance2015hardware, haider2017advancing}. The decision is being made whether a chip is tampered with a hardware Trojan by observing a mismatch between the observed and expected responses. Note that the accuracy of the detection process does not depend on the manufacturing process variations. However, the detection will be extremely difficult as it is practically impossible to detect all types of combinational Trojans. In addition, it is not feasible to trigger a sequential Trojan, as it requires to apply the same trigger pattern at the input $R$ times. 

Side-channel information, such as, power~\cite{Wei_2011}, temperature~\cite{Nowroz_2014}, delay~\cite{Jin_2008}, and radiation~\cite{he2017hardware} can be used to detect a hardware Trojan. The side-channel fingerprinting technique for detecting Analog/RF hardware Trojans in a wireless cryptographic IC has also been proposed~\cite{jin2010hardware, liu2015concurrent}. These detection methods rely on the availability of Trojan-free golden circuits for creating Trojan free signature. It can be very difficult to acquire a golden sample as all the chips may have Trojans. Path delay-based testing has limitations on detecting a hardware Trojan as long as the Trojan is inactive. Note that activating a Trojan is challenging as an adversary can select a hard to activate Trojan. In addition, the path delay does not depend on the size of the trigger circuit or the number of payloads attached with each trigger as the Trojan remains quiet during the logic testing. In addition, process and environmental variations may mask the side-channel leakage, when a Trojan circuitry is small. 

While dedicated towards hardware Trojan detection, researchers propose different measures to prevent a Trojan from being inserted into the design in the first place. These solutions involved characterization of ring-oscillator~\cite{rajendran2011design}, shadow registers~\cite{li2008speed}, and delay elements~\cite{ramdas2014slack} to detect the delay deviation caused by hardware Trojans. Reducing the rare signal in the circuitry is another proposed method for designers to reduce the risk of being implanted with a Trojan~\cite{salmani2012novel, zhou2014low}. Camouflage fill techniques~\cite{cocchi2014circuit, rajendran2013security} to create indistinguishable layouts for different gates by adding dummy contacts and connections can prevent the attacker from extracting a correct gate-level netlist of a circuit for Trojan insertion. Xiao \el proposed to fill all the unused spaces using filler cells so that an untrusted foundry cannot insert a Trojan~\cite{xiao2013bisa, shi2017obfuscated}. However, this direction still lacks any firm solution as more emphasis is observed in Trojan detection.  

Split Manufacturing can be an effective way to thwart Hardware Trojan insertion at an untrusted foundry. In split manufacturing, the production of ICs is carried out in two different foundries \cite{IAPRA2011}. The design is divided into two parts -- Front End of Line (FEOL) and Back End of Line (BEOL). An untrusted foundry is provided with the FEOL design, which contains partial information regarding the design that requires complex steps for fabricating and involves higher cost. Fabrication of BEOL does not incorporate complex fabrication steps and can be done by a smaller trusted foundry. The untrusted foundry sends the fabricated wafers directly to the smaller foundry for the complete fabrication. This way the untrusted foundry can be restricted to make any Trojan based modification as it does not have the complete information regarding the design. However, several attacks undermining the security achieved through split manufacturing have also been proposed in past~\cite{magana2017proximity, wang2017front, xu2019layout}. %Good Summary

Recent research contributions showed that machine learning and image processing can also be incorporated to detect hardware Trojans in the chip. Vashistha et al. presented Trojan scanner~\cite{vashistha2018trojan}, which uses a trusted GDSII layout (golden layout) and scanning electron microscope (SEM) images to identify the malicious modifications made in the netlist during the manufacturing of a circuit. A unique descriptor for each type of gate is prepared based on different features using computer vision algorithms along with a machine-learning model of a golden layout and SEM images of an IC under authentication. These descriptors, when compared to each other can detect any modifications either in the form of additional gates or modified gates which might raise the suspicion for a potential hardware Trojan. Moreover, Trojan scanner also presents the trade-off between the accuracy and SEM parameters. The authors demonstrated the effectiveness of the scheme using a smart card die as a test sample (generally manufactured with 90 \textit{nm} technology). It is yet to be validated its effectiveness of detection when a chip is fabricated using recent technology nodes (10 \textit{nm} and beyond).  

%{\color{red}
Research groups have also investigated vulnerabilities in the analog/RF front-end of a wireless device that can facilitate hardware Trojan attacks. Subramani \el proposed defense to distinguish between channel-induced and Trojan-induced impact on the legitimate signal through its traits~\cite{subramani2017ace}. Acknowledging excessive toggling activity on a victim wire of Trojans such as A2~\cite{yang2016a2}, Hou \el proposed to add on-chip monitors to measure switching activity on potential victim wires during an adjustable time period and raises an alarm if such activity goes above a certain threshold~\cite{hou2018r2d2}. This method can be effective for detecting A2 Trojan which comprises of a capacitor as a trigger. Reverse engineering based detection methods have also been proposed by sorting trace lengths to realize the minimum capacitance needed for such Trojans~\cite{guo2019capacitors}. However, it is possible to mitigate this requirement by an adversary, \eg using multiple layers/higher voltages to evade detection by this kind of approaches.%} 

Despite significant research have been performed on detecting hardware Trojans, we still lack efficient and accurate methods for modeling them and generating tests for their detection. Once the detection of hardware Trojans is ensured, an SoC designer can choose a SAT-resistant logic locking to prevent IC overproduction and IP piracy. 

\section{Conclusion}\label{conclusion}
In this paper, we have demonstrated the vulnerability of logic locking techniques through a set of tampering attacks with hardware Trojans. Three types of proposed \textit{TAAL attacks} can defeat any logic locking techniques that rely on storing the secret key in a tamper-proof memory. In \textit{T1 type TAAL Attack}, we showed how an adversary could extract the key from a locked netlist without knowing the details of the logic locking technique used to protect the circuit. For \textit{T2 type} and \textit{T3 type TAAL}  attacks, the complexity of detecting an attack has been improved. Only the attacker knows the specific values that can lead to key extraction, increasing the identification of a \textit{TAAL} attack. To launch a \textit{TAAL} attack, we develop models for combinational and sequential hardware Trojans. We also proposed an algorithm to design a hardware Trojan that cannot be detected by manufacturing tests. The results depict the range of Trojans selected by an adversary, which has a very high order of magnitude. Finally, we describe relevant detection and avoidance strategies for hardware Trojans to make logic locking secure. 

\section*{Acknowledgment}
This work was supported in part by the National Science Foundation under grant number CNS-1755733, and the United States Air Force/Air Force Materiel Command (USAF/AFMC) under grant AF-FA8650-19-1-1707. Any opinions, findings, and conclusions or recommendations expressed in this material are those of the authors and do not necessarily reflect the views of the NSF and USAF/AFMC.

\balance

\bibliographystyle{ACM-Reference-Format}
\bibliography{TODAES}


\begin{thebibliography}{107}

%%% the Web address.

\ifx \showCODEN    \undefined \def \showCODEN     #1{\unskip}     \fi
\ifx \showDOI      \undefined \def \showDOI       #1{#1}\fi
\ifx \showISBNx    \undefined \def \showISBNx     #1{\unskip}     \fi
\ifx \showISBNxiii \undefined \def \showISBNxiii  #1{\unskip}     \fi
\ifx \showISSN     \undefined \def \showISSN      #1{\unskip}     \fi
\ifx \showLCCN     \undefined \def \showLCCN      #1{\unskip}     \fi
\ifx \shownote     \undefined \def \shownote      #1{#1}          \fi
\ifx \showarticletitle \undefined \def \showarticletitle #1{#1}   \fi
\ifx \showURL      \undefined \def \showURL       {\relax}        \fi
% invisible to TeX
\providecommand\bibfield[2]{#2}
\providecommand\bibinfo[2]{#2}
\providecommand\natexlab[1]{#1}
\providecommand\showeprint[2][]{arXiv:#2}

\bibitem[\protect\citeauthoryear{Adee}{Adee}{2008}]%
        {adee2008hunt}
\bibfield{author}{\bibinfo{person}{Sally Adee}.}
  \bibinfo{year}{2008}\natexlab{}.
\newblock \showarticletitle{{{The Hunt for the Kill Switch}}}.
\newblock \bibinfo{journal}{\emph{IEEE Spectrum}} \bibinfo{volume}{45},
  \bibinfo{number}{5} (\bibinfo{year}{2008}), \bibinfo{pages}{34--39}.
\newblock


\bibitem[\protect\citeauthoryear{{Age Yeh}}{{Age Yeh}}{2012}]%
        {YehFabCost2012}
\bibfield{author}{\bibinfo{person}{{Age Yeh}}.}
  \bibinfo{year}{2012}\natexlab{}.
\newblock \bibinfo{title}{{Trends in the global IC design service market}}.
\newblock \bibinfo{howpublished}{DIGITIMES Research}.
\newblock


\bibitem[\protect\citeauthoryear{Agrawal, Baktir, Karakoyunlu, Rohatgi, and
  Sunar}{Agrawal et~al\mbox{.}}{2007}]%
        {agrawal2007trojan}
\bibfield{author}{\bibinfo{person}{Dakshi Agrawal}, \bibinfo{person}{Selcuk
  Baktir}, \bibinfo{person}{Deniz Karakoyunlu}, \bibinfo{person}{Pankaj
  Rohatgi}, {and} \bibinfo{person}{Berk Sunar}.}
  \bibinfo{year}{2007}\natexlab{}.
\newblock \showarticletitle{{{Trojan Detection Using IC Fingerprinting}}}. In
  \bibinfo{booktitle}{\emph{Proc. IEEE Symp. Security and Privacy (SP)}}.
  \bibinfo{pages}{296--310}.
\newblock


\bibitem[\protect\citeauthoryear{Alkabani and Koushanfar}{Alkabani and
  Koushanfar}{2007}]%
        {alkabani2007active}
\bibfield{author}{\bibinfo{person}{Yousra Alkabani} {and}
  \bibinfo{person}{Farinaz Koushanfar}.} \bibinfo{year}{2007}\natexlab{}.
\newblock \showarticletitle{{Active Hardware Metering for Intellectual Property
  Protection and Security.}}. In \bibinfo{booktitle}{\emph{USENIX security
  symposium}}. \bibinfo{pages}{291--306}.
\newblock


\bibitem[\protect\citeauthoryear{Alkabani, Koushanfar, and Potkonjak}{Alkabani
  et~al\mbox{.}}{2007}]%
        {alkabani2007remote}
\bibfield{author}{\bibinfo{person}{Yousra Alkabani}, \bibinfo{person}{Farinaz
  Koushanfar}, {and} \bibinfo{person}{Miodrag Potkonjak}.}
  \bibinfo{year}{2007}\natexlab{}.
\newblock \showarticletitle{{Remote activation of ICs for piracy prevention and
  digital right management}}. In \bibinfo{booktitle}{\emph{Proc. of IEEE/ACM
  int. conf. on Computer-aided design}}. \bibinfo{pages}{674--677}.
\newblock


\bibitem[\protect\citeauthoryear{Banga and Hsiao}{Banga and Hsiao}{2008}]%
        {banga2008region}
\bibfield{author}{\bibinfo{person}{Mainak Banga} {and}
  \bibinfo{person}{Michael~S Hsiao}.} \bibinfo{year}{2008}\natexlab{}.
\newblock \showarticletitle{{{A Region Based Approach for the Identification of
  Hardware Trojans}}}. In \bibinfo{booktitle}{\emph{Proc. IEEE Int. Workshop on
  Hardware-Oriented Security and Trust}}. \bibinfo{pages}{40--47}.
\newblock


\bibitem[\protect\citeauthoryear{Banga and Hsiao}{Banga and Hsiao}{2009}]%
        {banga2009novel}
\bibfield{author}{\bibinfo{person}{Mainak Banga} {and}
  \bibinfo{person}{Michael~S Hsiao}.} \bibinfo{year}{2009}\natexlab{}.
\newblock \showarticletitle{{A Novel Sustained Vector Technique for the
  Detection of Hardware Trojans}}. In \bibinfo{booktitle}{\emph{Proceedings of
  International Conference VLSI Design}}. \bibinfo{pages}{327--332}.
\newblock


\bibitem[\protect\citeauthoryear{Banga and Hsiao}{Banga and Hsiao}{2011}]%
        {banga2011odette}
\bibfield{author}{\bibinfo{person}{Mainak Banga} {and}
  \bibinfo{person}{Michael~S Hsiao}.} \bibinfo{year}{2011}\natexlab{}.
\newblock \showarticletitle{{{Odette: A Non-Scan Design-for-Test Methodology
  for Trojan Detection in ICs}}}. In \bibinfo{booktitle}{\emph{Proc. IEEE Int.
  Symp. Hardware-Oriented Security and Trust}}. \bibinfo{pages}{18--23}.
\newblock


\bibitem[\protect\citeauthoryear{Baumgarten, Tyagi, and Zambreno}{Baumgarten
  et~al\mbox{.}}{2010}]%
        {baumgarten2010preventing}
\bibfield{author}{\bibinfo{person}{Alex Baumgarten}, \bibinfo{person}{Akhilesh
  Tyagi}, {and} \bibinfo{person}{Joseph Zambreno}.}
  \bibinfo{year}{2010}\natexlab{}.
\newblock \showarticletitle{{Preventing IC piracy using reconfigurable logic
  barriers}}.
\newblock \bibinfo{journal}{\emph{IEEE Design \& Test of Computers}}
  \bibinfo{volume}{27}, \bibinfo{number}{1} (\bibinfo{year}{2010}),
  \bibinfo{pages}{66--75}.
\newblock


\bibitem[\protect\citeauthoryear{Bhunia, Hsiao, Banga, and Narasimhan}{Bhunia
  et~al\mbox{.}}{2014}]%
        {bhunia2014hardware}
\bibfield{author}{\bibinfo{person}{Swarup Bhunia}, \bibinfo{person}{Michael~S
  Hsiao}, \bibinfo{person}{Mainak Banga}, {and} \bibinfo{person}{Seetharam
  Narasimhan}.} \bibinfo{year}{2014}\natexlab{}.
\newblock \showarticletitle{{{Hardware Trojan Attacks: Threat Analysis and
  Countermeasures}}}.
\newblock \bibinfo{journal}{\emph{Proc. IEEE}} \bibinfo{volume}{102},
  \bibinfo{number}{8} (\bibinfo{year}{2014}), \bibinfo{pages}{1229--1247}.
\newblock


\bibitem[\protect\citeauthoryear{Bhunia and Tehranipoor}{Bhunia and
  Tehranipoor}{2018}]%
        {bhunia2018hardware}
\bibfield{author}{\bibinfo{person}{Swarup Bhunia} {and} \bibinfo{person}{Mark
  Tehranipoor}.} \bibinfo{year}{2018}\natexlab{}.
\newblock \bibinfo{booktitle}{\emph{{Hardware Security: A Hands-on Learning
  Approach}}}.
\newblock \bibinfo{publisher}{Morgan Kaufmann}.
\newblock


\bibitem[\protect\citeauthoryear{Bryan}{Bryan}{1985}]%
        {bryan1985iscas}
\bibfield{author}{\bibinfo{person}{David Bryan}.}
  \bibinfo{year}{1985}\natexlab{}.
\newblock \showarticletitle{{The ISCAS'85 benchmark circuits and netlist
  format}}.
\newblock \bibinfo{journal}{\emph{North Carolina State University}}
  \bibinfo{volume}{25} (\bibinfo{year}{1985}).
\newblock


\bibitem[\protect\citeauthoryear{Bushnell and Agrawal}{Bushnell and
  Agrawal}{2004}]%
        {bushnell2004essentials}
\bibfield{author}{\bibinfo{person}{Michael Bushnell} {and}
  \bibinfo{person}{Vishwani Agrawal}.} \bibinfo{year}{2004}\natexlab{}.
\newblock \bibinfo{booktitle}{\emph{{Essentials of electronic testing for
  digital, memory and mixed-signal VLSI circuits}}}. Vol.~\bibinfo{volume}{17}.
\newblock \bibinfo{publisher}{Springer Science \& Business Media}.
\newblock


\bibitem[\protect\citeauthoryear{Castillo, Meyer-Baese, Garc{\'\i}a, Parrilla,
  and Lloris}{Castillo et~al\mbox{.}}{2007}]%
        {castillo2007ipp}
\bibfield{author}{\bibinfo{person}{Encarnacin Castillo}, \bibinfo{person}{Uwe
  Meyer-Baese}, \bibinfo{person}{Antonio Garc{\'\i}a}, \bibinfo{person}{Luis
  Parrilla}, {and} \bibinfo{person}{Antonio Lloris}.}
  \bibinfo{year}{2007}\natexlab{}.
\newblock \showarticletitle{{IPP@ HDL: efficient intellectual property
  protection scheme for IP cores}}.
\newblock \bibinfo{journal}{\emph{IEEE Transactions on Very Large Scale
  Integration (VLSI) Systems}} \bibinfo{volume}{15}, \bibinfo{number}{5}
  (\bibinfo{year}{2007}), \bibinfo{pages}{578--591}.
\newblock


\bibitem[\protect\citeauthoryear{Chakraborty and Bhunia}{Chakraborty and
  Bhunia}{2008}]%
        {chakraborty2008hardware}
\bibfield{author}{\bibinfo{person}{Rajat~Subhra Chakraborty} {and}
  \bibinfo{person}{Swarup Bhunia}.} \bibinfo{year}{2008}\natexlab{}.
\newblock \showarticletitle{Hardware protection and authentication through
  netlist level obfuscation}. In \bibinfo{booktitle}{\emph{Proc. of IEEE/ACM
  International Conference on Computer-Aided Design}}.
  \bibinfo{pages}{674--677}.
\newblock


\bibitem[\protect\citeauthoryear{Chakraborty and Bhunia}{Chakraborty and
  Bhunia}{2009}]%
        {chakraborty2009security}
\bibfield{author}{\bibinfo{person}{Rajat~Subhra Chakraborty} {and}
  \bibinfo{person}{Swarup Bhunia}.} \bibinfo{year}{2009}\natexlab{}.
\newblock \showarticletitle{{{Security Against Hardware Trojan Through a Novel
  Application of Design Obfuscation}}}. In \bibinfo{booktitle}{\emph{Proc. Int.
  Conf. Computer-Aided Design}}. \bibinfo{pages}{113--116}.
\newblock


\bibitem[\protect\citeauthoryear{Chakraborty, Wolff, Paul, Papachristou, and
  Bhunia}{Chakraborty et~al\mbox{.}}{2009}]%
        {chakraborty2009mero}
\bibfield{author}{\bibinfo{person}{Rajat~Subhra Chakraborty},
  \bibinfo{person}{Francis~G Wolff}, \bibinfo{person}{Somnath Paul},
  \bibinfo{person}{Christos~A Papachristou}, {and} \bibinfo{person}{Swarup
  Bhunia}.} \bibinfo{year}{2009}\natexlab{}.
\newblock \showarticletitle{{{MERO: A Statistical Approach for Hardware Trojan
  Detection}}}. In \bibinfo{booktitle}{\emph{Proc. International Workshop on
  Cryptographic Hardware and Embedded Systems (CHES)}}.
  \bibinfo{pages}{396--410}.
\newblock


\bibitem[\protect\citeauthoryear{Chang, Bakkaloglu, and Ozev}{Chang
  et~al\mbox{.}}{2015}]%
        {chang2015enabling}
\bibfield{author}{\bibinfo{person}{Doohwang Chang}, \bibinfo{person}{Bertan
  Bakkaloglu}, {and} \bibinfo{person}{Sule Ozev}.}
  \bibinfo{year}{2015}\natexlab{}.
\newblock \showarticletitle{{Enabling unauthorized RF transmission below noise
  floor with no detectable impact on primary communication performance}}. In
  \bibinfo{booktitle}{\emph{VLSI Test Symposium (VTS)}}. \bibinfo{pages}{1--4}.
\newblock


\bibitem[\protect\citeauthoryear{Charbon}{Charbon}{1998}]%
        {charbon1998hierarchical}
\bibfield{author}{\bibinfo{person}{Edoardo Charbon}.}
  \bibinfo{year}{1998}\natexlab{}.
\newblock \showarticletitle{{Hierarchical watermarking in IC design}}. In
  \bibinfo{booktitle}{\emph{Proc. of the Custom Integrated Circuits}}.
  \bibinfo{pages}{295--298}.
\newblock


\bibitem[\protect\citeauthoryear{Cocchi, Baukus, Chow, and Wang}{Cocchi
  et~al\mbox{.}}{2014}]%
        {cocchi2014circuit}
\bibfield{author}{\bibinfo{person}{Ronald~P Cocchi}, \bibinfo{person}{James~P
  Baukus}, \bibinfo{person}{Lap~Wai Chow}, {and} \bibinfo{person}{Bryan~J
  Wang}.} \bibinfo{year}{2014}\natexlab{}.
\newblock \showarticletitle{{Circuit camouflage integration for hardware IP
  protection}}. In \bibinfo{booktitle}{\emph{Proceedings of Annual Design
  Automation Conference}}. \bibinfo{pages}{1--5}.
\newblock


\bibitem[\protect\citeauthoryear{Contreras, Rahman, and Tehranipoor}{Contreras
  et~al\mbox{.}}{2013}]%
        {contreras2013secure}
\bibfield{author}{\bibinfo{person}{Gustavo~K Contreras},
  \bibinfo{person}{Md~Tauhidur Rahman}, {and} \bibinfo{person}{Mohammad
  Tehranipoor}.} \bibinfo{year}{2013}\natexlab{}.
\newblock \showarticletitle{{Secure split-test for preventing IC piracy by
  untrusted foundry and assembly}}. In \bibinfo{booktitle}{\emph{IEEE
  International symposium on defect and fault tolerance in VLSI and
  nanotechnology systems (DFTS)}}. \bibinfo{pages}{196--203}.
\newblock


\bibitem[\protect\citeauthoryear{Davidson}{Davidson}{1999}]%
        {davidson1999itc}
\bibfield{author}{\bibinfo{person}{Scott Davidson}.}
  \bibinfo{year}{1999}\natexlab{}.
\newblock \showarticletitle{{ITC'99 benchmark circuits-preliminary results}}.
  In \bibinfo{booktitle}{\emph{International Test Conference 1999. Proceedings
  (IEEE Cat. No. 99CH37034)}}. IEEE Computer Society,
  \bibinfo{pages}{1125--1125}.
\newblock


\bibitem[\protect\citeauthoryear{{DC Ultra: Concurrent Timing, Area, Power, and
  Test Optimization}}{{DC Ultra: Concurrent Timing, Area, Power, and Test
  Optimization}}{2019}]%
        {SynopsysDC}
\bibfield{author}{\bibinfo{person}{{DC Ultra: Concurrent Timing, Area, Power,
  and Test Optimization}}.} \bibinfo{year}{2019}\natexlab{}.
\newblock \bibinfo{title}{[Online] Available at:
  https://www.synopsys.com/implementation-and-signoff/rtl-synthesis-test/dc-ultra.html}.
\newblock
\newblock


\bibitem[\protect\citeauthoryear{Guin, Shi, Forte, and Tehranipoor}{Guin
  et~al\mbox{.}}{2016}]%
        {guin2016fortis}
\bibfield{author}{\bibinfo{person}{Ujjwal Guin}, \bibinfo{person}{Qihang Shi},
  \bibinfo{person}{Domenic Forte}, {and} \bibinfo{person}{Mark~M Tehranipoor}.}
  \bibinfo{year}{2016}\natexlab{}.
\newblock \showarticletitle{{FORTIS: a comprehensive solution for establishing
  forward trust for protecting IPs and ICs}}.
\newblock \bibinfo{journal}{\emph{ACM Transactions on Design Automation of
  Electronic Systems (TODAES)}} \bibinfo{volume}{21}, \bibinfo{number}{4}
  (\bibinfo{year}{2016}), \bibinfo{pages}{63}.
\newblock


\bibitem[\protect\citeauthoryear{Guin, Zhou, and Singh}{Guin
  et~al\mbox{.}}{2017}]%
        {GuinVTS2017}
\bibfield{author}{\bibinfo{person}{U. Guin}, \bibinfo{person}{Ziqi Zhou}, {and}
  \bibinfo{person}{A. Singh}.} \bibinfo{year}{2017}\natexlab{}.
\newblock \showarticletitle{{A novel design-for-security (DFS) architecture to
  prevent unauthorized IC overproduction}}. In \bibinfo{booktitle}{\emph{Proc.
  of the IEEE VLSI Test Symposium (VTS)}}. \bibinfo{pages}{1--6}.
\newblock


\bibitem[\protect\citeauthoryear{Guin, Zhou, and Singh}{Guin
  et~al\mbox{.}}{2018}]%
        {GuinTVLSI2018}
\bibfield{author}{\bibinfo{person}{Ujjwal Guin}, \bibinfo{person}{Ziqi Zhou},
  {and} \bibinfo{person}{Adit Singh}.} \bibinfo{year}{2018}\natexlab{}.
\newblock \showarticletitle{{Robust design-for-security architecture for
  enabling trust in IC manufacturing and test}}.
\newblock \bibinfo{journal}{\emph{Trans. on Very Large Scale Integration (VLSI)
  Systems}} \bibinfo{volume}{26}, \bibinfo{number}{5} (\bibinfo{year}{2018}),
  \bibinfo{pages}{818--830}.
\newblock


\bibitem[\protect\citeauthoryear{Guo, Zhu, Jin, and Zhang}{Guo
  et~al\mbox{.}}{2019}]%
        {guo2019capacitors}
\bibfield{author}{\bibinfo{person}{Xiaolong Guo}, \bibinfo{person}{Huifeng
  Zhu}, \bibinfo{person}{Yier Jin}, {and} \bibinfo{person}{Xuan Zhang}.}
  \bibinfo{year}{2019}\natexlab{}.
\newblock \showarticletitle{{When Capacitors Attack: Formal Method Driven
  Design and Detection of Charge-Domain Trojans}}. In
  \bibinfo{booktitle}{\emph{Design, Automation \& Test in Europe Conf. \&
  Exhibition (DATE)}}.
\newblock


\bibitem[\protect\citeauthoryear{Haider, Jin, Ahmad, Shila, Khan, and van
  Dijk}{Haider et~al\mbox{.}}{2017}]%
        {haider2017advancing}
\bibfield{author}{\bibinfo{person}{Syed~Kamran Haider},
  \bibinfo{person}{Chenglu Jin}, \bibinfo{person}{Masab Ahmad},
  \bibinfo{person}{Devu Shila}, \bibinfo{person}{Omer Khan}, {and}
  \bibinfo{person}{Marten van Dijk}.} \bibinfo{year}{2017}\natexlab{}.
\newblock \showarticletitle{{{Advancing the State-of-the-Art in Hardware
  Trojans Detection}}}.
\newblock \bibinfo{journal}{\emph{IEEE Transactions on Dependable and Secure
  Computing}} (\bibinfo{year}{2017}).
\newblock


\bibitem[\protect\citeauthoryear{He, Zhao, Guo, and Jin}{He
  et~al\mbox{.}}{2017}]%
        {he2017hardware}
\bibfield{author}{\bibinfo{person}{Jiaji He}, \bibinfo{person}{Yiqiang Zhao},
  \bibinfo{person}{Xiaolong Guo}, {and} \bibinfo{person}{Yier Jin}.}
  \bibinfo{year}{2017}\natexlab{}.
\newblock \showarticletitle{{{Hardware Trojan Detection Through Chip-Free
  Electromagnetic Side-Channel Statistical Analysis}}}.
\newblock \bibinfo{journal}{\emph{IEEE Trans. Very Large Scale Integration
  Sys.}} \bibinfo{volume}{25}, \bibinfo{number}{10} (\bibinfo{year}{2017}),
  \bibinfo{pages}{2939--2948}.
\newblock


\bibitem[\protect\citeauthoryear{Hou, He, Shamsi, Jin, Wu, and Wu}{Hou
  et~al\mbox{.}}{2018}]%
        {hou2018r2d2}
\bibfield{author}{\bibinfo{person}{Yumin Hou}, \bibinfo{person}{Hu He},
  \bibinfo{person}{Kaveh Shamsi}, \bibinfo{person}{Yier Jin},
  \bibinfo{person}{Dong Wu}, {and} \bibinfo{person}{Huaqiang Wu}.}
  \bibinfo{year}{2018}\natexlab{}.
\newblock \showarticletitle{{R2d2: Runtime reassurance and detection of A2
  Trojan}}. In \bibinfo{booktitle}{\emph{International Symposium on Hardware
  Oriented Security and Trust (HOST)}}. \bibinfo{pages}{195--200}.
\newblock


\bibitem[\protect\citeauthoryear{Huang and Lach}{Huang and Lach}{2008}]%
        {huang2008ic}
\bibfield{author}{\bibinfo{person}{Jiawei Huang} {and} \bibinfo{person}{John
  Lach}.} \bibinfo{year}{2008}\natexlab{}.
\newblock \showarticletitle{{IC activation and user authentication for
  security-sensitive systems}}. In \bibinfo{booktitle}{\emph{IEEE International
  Workshop on Hardware-Oriented Security and Trust}}. \bibinfo{pages}{76--80}.
\newblock


\bibitem[\protect\citeauthoryear{{Intelligence Advanced Research Projects
  Activity}}{{Intelligence Advanced Research Projects Activity}}{2011}]%
        {IAPRA2011}
\bibfield{author}{\bibinfo{person}{{Intelligence Advanced Research Projects
  Activity}}.} \bibinfo{year}{2011}\natexlab{}.
\newblock \bibinfo{title}{{Trusted Integrated Chips (TIC) Program}}.
\newblock
\newblock


\bibitem[\protect\citeauthoryear{Jarvis and McIntyre}{Jarvis and
  McIntyre}{2007}]%
        {jarvis2007split}
\bibfield{author}{\bibinfo{person}{Richard~Wayne Jarvis} {and}
  \bibinfo{person}{Michael~G McIntyre}.} \bibinfo{year}{2007}\natexlab{}.
\newblock \bibinfo{title}{Split manufacturing method for advanced semiconductor
  circuits}.
\newblock
\newblock
\newblock
\shownote{US Patent 7,195,931.}


\bibitem[\protect\citeauthoryear{Jin and Makris}{Jin and Makris}{2008}]%
        {Jin_2008}
\bibfield{author}{\bibinfo{person}{Y. Jin} {and} \bibinfo{person}{Y. Makris}.}
  \bibinfo{year}{2008}\natexlab{}.
\newblock \showarticletitle{{{Hardware Trojan Detection Using Path Delay
  Fingerprint}}}. In \bibinfo{booktitle}{\emph{Proc. HOST}}.
  \bibinfo{pages}{51–--57}.
\newblock


\bibitem[\protect\citeauthoryear{Jin and Makris}{Jin and Makris}{2010}]%
        {jin2010hardware}
\bibfield{author}{\bibinfo{person}{Yier Jin} {and} \bibinfo{person}{Yiorgos
  Makris}.} \bibinfo{year}{2010}\natexlab{}.
\newblock \showarticletitle{{Hardware Trojans in wireless cryptographic ICs}}.
\newblock \bibinfo{journal}{\emph{IEEE Design \& Test of Computers}}
  (\bibinfo{year}{2010}), \bibinfo{pages}{26--35}.
\newblock


\bibitem[\protect\citeauthoryear{Kahng, Lach, Mangione-Smith, Mantik, Markov,
  Potkonjak, Tucker, Wang, and Wolfe}{Kahng et~al\mbox{.}}{2001}]%
        {kahng2001constraint}
\bibfield{author}{\bibinfo{person}{Andrew~B Kahng}, \bibinfo{person}{John
  Lach}, \bibinfo{person}{William~H Mangione-Smith}, \bibinfo{person}{Stefanus
  Mantik}, \bibinfo{person}{Igor~L Markov}, \bibinfo{person}{Miodrag
  Potkonjak}, \bibinfo{person}{Paul Tucker}, \bibinfo{person}{Huijuan Wang},
  {and} \bibinfo{person}{Gregory Wolfe}.} \bibinfo{year}{2001}\natexlab{}.
\newblock \showarticletitle{{Constraint-based watermarking techniques for
  design IP protection}}.
\newblock \bibinfo{journal}{\emph{IEEE Transactions on Computer-Aided Design of
  Integrated Circuits and Systems}} \bibinfo{volume}{20}, \bibinfo{number}{10}
  (\bibinfo{year}{2001}), \bibinfo{pages}{1236--1252}.
\newblock


\bibitem[\protect\citeauthoryear{Karri, Rajendran, Rosenfeld, and
  Tehranipoor}{Karri et~al\mbox{.}}{2010}]%
        {karri2010trustworthy}
\bibfield{author}{\bibinfo{person}{Ramesh Karri}, \bibinfo{person}{Jeyavijayan
  Rajendran}, \bibinfo{person}{Kurt Rosenfeld}, {and} \bibinfo{person}{Mohammad
  Tehranipoor}.} \bibinfo{year}{2010}\natexlab{}.
\newblock \showarticletitle{{{Trustworthy Hardware: Identifying and Classifying
  Hardware Trojans}}}.
\newblock \bibinfo{journal}{\emph{Computer}} \bibinfo{volume}{43},
  \bibinfo{number}{10} (\bibinfo{year}{2010}), \bibinfo{pages}{39--46}.
\newblock


\bibitem[\protect\citeauthoryear{Khaleghi, Da~Zhao, and Rao}{Khaleghi
  et~al\mbox{.}}{2015}]%
        {khaleghi2015ic}
\bibfield{author}{\bibinfo{person}{Soroush Khaleghi}, \bibinfo{person}{Kai
  Da~Zhao}, {and} \bibinfo{person}{Wenjing Rao}.}
  \bibinfo{year}{2015}\natexlab{}.
\newblock \showarticletitle{{IC piracy prevention via design withholding and
  entanglement}}. In \bibinfo{booktitle}{\emph{Asia and South Pacific Design
  Automation Conference}}. \bibinfo{pages}{821--826}.
\newblock


\bibitem[\protect\citeauthoryear{Kison, Awad, Fyrbiak, and Paar}{Kison
  et~al\mbox{.}}{2019}]%
        {kison2019security}
\bibfield{author}{\bibinfo{person}{Christian Kison},
  \bibinfo{person}{Omar~Mohamed Awad}, \bibinfo{person}{Marc Fyrbiak}, {and}
  \bibinfo{person}{Christof Paar}.} \bibinfo{year}{2019}\natexlab{}.
\newblock \showarticletitle{Security Implications of Intentional Capacitive
  Crosstalk}.
\newblock \bibinfo{journal}{\emph{Trans. on Information Forensics and
  Security}} (\bibinfo{year}{2019}).
\newblock


\bibitem[\protect\citeauthoryear{Koushanfar and Qu}{Koushanfar and Qu}{2001}]%
        {koushanfar2001hardware}
\bibfield{author}{\bibinfo{person}{Farinaz Koushanfar} {and}
  \bibinfo{person}{Gang Qu}.} \bibinfo{year}{2001}\natexlab{}.
\newblock \showarticletitle{Hardware metering}. In
  \bibinfo{booktitle}{\emph{Proceedings of Design Automation Conference}}.
  \bibinfo{pages}{490--493}.
\newblock


\bibitem[\protect\citeauthoryear{Lee and Touba}{Lee and Touba}{2015}]%
        {lee2015improving}
\bibfield{author}{\bibinfo{person}{Yu-Wei Lee} {and} \bibinfo{person}{Nur~A
  Touba}.} \bibinfo{year}{2015}\natexlab{}.
\newblock \showarticletitle{Improving logic obfuscation via logic cone
  analysis}. In \bibinfo{booktitle}{\emph{Latin-American Test Symposium
  (LATS)}}. \bibinfo{pages}{1--6}.
\newblock


\bibitem[\protect\citeauthoryear{Lesperance, Kulkarni, and Cheng}{Lesperance
  et~al\mbox{.}}{2015}]%
        {lesperance2015hardware}
\bibfield{author}{\bibinfo{person}{Nicole Lesperance},
  \bibinfo{person}{Shrikant Kulkarni}, {and} \bibinfo{person}{Kwang-Ting
  Cheng}.} \bibinfo{year}{2015}\natexlab{}.
\newblock \showarticletitle{{{Hardware Trojan Detection Using Exhaustive
  Testing of k-bit Subspaces}}}. In \bibinfo{booktitle}{\emph{Proc. of Asia and
  South Pacific Design Automation Conf. (ASP-DAC)}}. \bibinfo{pages}{755--760}.
\newblock


\bibitem[\protect\citeauthoryear{Li and Lach}{Li and Lach}{2008}]%
        {li2008speed}
\bibfield{author}{\bibinfo{person}{Jie Li} {and} \bibinfo{person}{John Lach}.}
  \bibinfo{year}{2008}\natexlab{}.
\newblock \showarticletitle{{At-speed delay characterization for IC
  authentication and Trojan horse detection}}. In
  \bibinfo{booktitle}{\emph{IEEE International Workshop on Hardware-Oriented
  Security and Trust}}. \bibinfo{pages}{8--14}.
\newblock


\bibitem[\protect\citeauthoryear{Liu and Wang}{Liu and Wang}{2014}]%
        {liu2014embedded}
\bibfield{author}{\bibinfo{person}{Bao Liu} {and} \bibinfo{person}{Brandon
  Wang}.} \bibinfo{year}{2014}\natexlab{}.
\newblock \showarticletitle{{Embedded reconfigurable logic for ASIC design
  obfuscation against supply chain attacks}}. In
  \bibinfo{booktitle}{\emph{{Design, Automation \& Test in Europe Conference \&
  Exhibition (DATE)}}}. \bibinfo{pages}{1--6}.
\newblock


\bibitem[\protect\citeauthoryear{Liu, Huang, and Makris}{Liu
  et~al\mbox{.}}{2014}]%
        {liu2014hardware}
\bibfield{author}{\bibinfo{person}{Yu Liu}, \bibinfo{person}{Ke Huang}, {and}
  \bibinfo{person}{Yiorgos Makris}.} \bibinfo{year}{2014}\natexlab{}.
\newblock \showarticletitle{{{Hardware Trojan Detection Through Golden
  Chip-Free Statistical Side-Channel Fingerprinting}}}. In
  \bibinfo{booktitle}{\emph{Proc. of Design Automation Conference}}.
\newblock


\bibitem[\protect\citeauthoryear{Liu, Volanis, Huang, and Makris}{Liu
  et~al\mbox{.}}{2015}]%
        {liu2015concurrent}
\bibfield{author}{\bibinfo{person}{Yu Liu}, \bibinfo{person}{Georgios Volanis},
  \bibinfo{person}{Ke Huang}, {and} \bibinfo{person}{Yiorgos Makris}.}
  \bibinfo{year}{2015}\natexlab{}.
\newblock \showarticletitle{Concurrent hardware Trojan detection in wireless
  cryptographic ICs}. In \bibinfo{booktitle}{\emph{International Test
  Conference (ITC)}}. \bibinfo{pages}{1--8}.
\newblock


\bibitem[\protect\citeauthoryear{Maga{\~n}a, Shi, Melchert, and
  Davoodi}{Maga{\~n}a et~al\mbox{.}}{2017}]%
        {magana2017proximity}
\bibfield{author}{\bibinfo{person}{Jonathon Maga{\~n}a},
  \bibinfo{person}{Daohang Shi}, \bibinfo{person}{Jackson Melchert}, {and}
  \bibinfo{person}{Azadeh Davoodi}.} \bibinfo{year}{2017}\natexlab{}.
\newblock \showarticletitle{Are proximity attacks a threat to the security of
  split manufacturing of integrated circuits?}
\newblock \bibinfo{journal}{\emph{Trans. on Very Large Scale Integration (VLSI)
  Systems}} (\bibinfo{year}{2017}), \bibinfo{pages}{3406--3419}.
\newblock


\bibitem[\protect\citeauthoryear{Nagarajan, Khan, and Ghosh}{Nagarajan
  et~al\mbox{.}}{2019}]%
        {nagarajan2019entt}
\bibfield{author}{\bibinfo{person}{Karthikeyan Nagarajan},
  \bibinfo{person}{Mohammad Nasim~Imtiaz Khan}, {and} \bibinfo{person}{Swaroop
  Ghosh}.} \bibinfo{year}{2019}\natexlab{}.
\newblock \showarticletitle{{ENTT: A family of emerging NVM-based trojan
  triggers}}. In \bibinfo{booktitle}{\emph{International Symposium on Hardware
  Oriented Security and Trust (HOST)}}. \bibinfo{pages}{51--60}.
\newblock


\bibitem[\protect\citeauthoryear{Ngo, Bhasin, Danger, Guilley, and Najm}{Ngo
  et~al\mbox{.}}{2015}]%
        {ngo2015linear}
\bibfield{author}{\bibinfo{person}{Xuan~Thuy Ngo}, \bibinfo{person}{Shivam
  Bhasin}, \bibinfo{person}{Jean-Luc Danger}, \bibinfo{person}{Sylvain
  Guilley}, {and} \bibinfo{person}{Zakaria Najm}.}
  \bibinfo{year}{2015}\natexlab{}.
\newblock \showarticletitle{{{Linear Complementary Dual Code Improvement to
  Strengthen Encoded Circuit Against Hardware Trojan Horses}}}. In
  \bibinfo{booktitle}{\emph{Proc. IEEE Int. Symp. Hardware Oriented Security
  and Trust}}. \bibinfo{pages}{82--87}.
\newblock


\bibitem[\protect\citeauthoryear{nm~Generic Library~for Teaching
  IC~Design.}{nm~Generic Library~for Teaching IC~Design.}{2019}]%
        {Synopsyslibrary}
\bibfield{author}{\bibinfo{person}{Synopsys~32/28 nm~Generic Library~for
  Teaching IC~Design.}} \bibinfo{year}{Accessed 2019}\natexlab{}.
\newblock \bibinfo{title}{{[online] Available at:
  https://www.synopsys.com/community/universityprogram/teaching-resources.html}}.
\newblock
\newblock


\bibitem[\protect\citeauthoryear{Nowroz, Hu, Koushanfar, and Reda}{Nowroz
  et~al\mbox{.}}{2014}]%
        {Nowroz_2014}
\bibfield{author}{\bibinfo{person}{Abdullah~Nazma Nowroz},
  \bibinfo{person}{Kangqiao Hu}, \bibinfo{person}{Farinaz Koushanfar}, {and}
  \bibinfo{person}{Sherief Reda}.} \bibinfo{year}{2014}\natexlab{}.
\newblock \showarticletitle{{{Novel Techniques for High-Sensitivity Hardware
  Trojan Detection Using Thermal and Power Maps}}}.
\newblock \bibinfo{journal}{\emph{Trans. Computer-Aided Design of Integrated
  Circuits and Systems}} \bibinfo{volume}{33}, \bibinfo{number}{12}
  (\bibinfo{year}{2014}), \bibinfo{pages}{1792--1805}.
\newblock


\bibitem[\protect\citeauthoryear{Plaza and Markov}{Plaza and Markov}{2015}]%
        {plaza2015solving}
\bibfield{author}{\bibinfo{person}{Stephen~M Plaza} {and}
  \bibinfo{person}{Igor~L Markov}.} \bibinfo{year}{2015}\natexlab{}.
\newblock \showarticletitle{{Solving the third-shift problem in IC piracy with
  test-aware logic locking}}.
\newblock \bibinfo{journal}{\emph{Trans. on Computer-Aided Design of Integrated
  Circuits and Systems}} \bibinfo{volume}{34}, \bibinfo{number}{6}
  (\bibinfo{year}{2015}), \bibinfo{pages}{961--971}.
\newblock


\bibitem[\protect\citeauthoryear{Qu and Potkonjak}{Qu and Potkonjak}{2007}]%
        {qu2007intellectual}
\bibfield{author}{\bibinfo{person}{Gang Qu} {and} \bibinfo{person}{Miodrag
  Potkonjak}.} \bibinfo{year}{2007}\natexlab{}.
\newblock \bibinfo{booktitle}{\emph{Intellectual property protection in VLSI
  designs: theory and practice}}.
\newblock \bibinfo{publisher}{Springer Science \& Business Media}.
\newblock


\bibitem[\protect\citeauthoryear{Rad, Plusquellic, and Tehranipoor}{Rad
  et~al\mbox{.}}{2008}]%
        {rad2008sensitivity}
\bibfield{author}{\bibinfo{person}{Reza Rad}, \bibinfo{person}{Jim
  Plusquellic}, {and} \bibinfo{person}{Mohammad Tehranipoor}.}
  \bibinfo{year}{2008}\natexlab{}.
\newblock \showarticletitle{{{Sensitivity Analysis to Hardware Trojans Using
  Power Supply Transient Signals}}}. In \bibinfo{booktitle}{\emph{Proc. IEEE
  Int. Workshop on Hardware-Oriented Security and Trust}}.
  \bibinfo{pages}{3--7}.
\newblock


\bibitem[\protect\citeauthoryear{Rahman, Tajik, Rahman, Tehranipoor, and
  Asadizanjani}{Rahman et~al\mbox{.}}{2019}]%
        {rahman2019key}
\bibfield{author}{\bibinfo{person}{MT Rahman}, \bibinfo{person}{S Tajik},
  \bibinfo{person}{MS Rahman}, \bibinfo{person}{M Tehranipoor}, {and}
  \bibinfo{person}{N Asadizanjani}.} \bibinfo{year}{2019}\natexlab{}.
\newblock \bibinfo{booktitle}{\emph{{The key is left under the mat: On the
  inappropriate security assumption of logic locking schemes}}}.
\newblock \bibinfo{type}{{T}echnical {R}eport}.
\newblock


\bibitem[\protect\citeauthoryear{Rahman, Forte, Shi, Contreras, and
  Tehranipoor}{Rahman et~al\mbox{.}}{2014}]%
        {rahman2014csst}
\bibfield{author}{\bibinfo{person}{Md~Tauhidur Rahman},
  \bibinfo{person}{Domenic Forte}, \bibinfo{person}{Quihang Shi},
  \bibinfo{person}{Gustavo~K Contreras}, {and} \bibinfo{person}{Mohammad
  Tehranipoor}.} \bibinfo{year}{2014}\natexlab{}.
\newblock \showarticletitle{{CSST: an efficient secure split-test for
  preventing IC piracy}}. In \bibinfo{booktitle}{\emph{IEEE 23rd North Atlantic
  Test Workshop}}. \bibinfo{pages}{43--47}.
\newblock


\bibitem[\protect\citeauthoryear{Rajendran, Jyothi, Sinanoglu, and
  Karri}{Rajendran et~al\mbox{.}}{2011}]%
        {rajendran2011design}
\bibfield{author}{\bibinfo{person}{Jeyavijayan Rajendran},
  \bibinfo{person}{Vinayaka Jyothi}, \bibinfo{person}{Ozgur Sinanoglu}, {and}
  \bibinfo{person}{Ramesh Karri}.} \bibinfo{year}{2011}\natexlab{}.
\newblock \showarticletitle{{{Design and Analysis of Ring Oscillator Based
  Design-for-Trust Technique}}}. In \bibinfo{booktitle}{\emph{Proc. of VLSI
  Test Symp.}} \bibinfo{pages}{105--110}.
\newblock


\bibitem[\protect\citeauthoryear{Rajendran, Pino, Sinanoglu, and
  Karri}{Rajendran et~al\mbox{.}}{2012}]%
        {rajendran2012security}
\bibfield{author}{\bibinfo{person}{Jeyavijayan Rajendran},
  \bibinfo{person}{Youngok Pino}, \bibinfo{person}{Ozgur Sinanoglu}, {and}
  \bibinfo{person}{Ramesh Karri}.} \bibinfo{year}{2012}\natexlab{}.
\newblock \showarticletitle{Security analysis of logic obfuscation}. In
  \bibinfo{booktitle}{\emph{Proceedings of Annual Design Automation
  Conference}}. \bibinfo{pages}{83--89}.
\newblock


\bibitem[\protect\citeauthoryear{Rajendran, Sam, Sinanoglu, and
  Karri}{Rajendran et~al\mbox{.}}{2013a}]%
        {rajendran2013security}
\bibfield{author}{\bibinfo{person}{Jeyavijayan Rajendran},
  \bibinfo{person}{Michael Sam}, \bibinfo{person}{Ozgur Sinanoglu}, {and}
  \bibinfo{person}{Ramesh Karri}.} \bibinfo{year}{2013}\natexlab{a}.
\newblock \showarticletitle{Security analysis of integrated circuit
  camouflaging}. In \bibinfo{booktitle}{\emph{Proc. of ACM SIGSAC conference on
  Computer \& communications security}}. \bibinfo{pages}{709--720}.
\newblock


\bibitem[\protect\citeauthoryear{Rajendran, Zhang, Zhang, Rose, Pino,
  Sinanoglu, and Karri}{Rajendran et~al\mbox{.}}{2015}]%
        {rajendran2015fault}
\bibfield{author}{\bibinfo{person}{Jeyavijayan Rajendran},
  \bibinfo{person}{Huan Zhang}, \bibinfo{person}{Chi Zhang},
  \bibinfo{person}{Garrett~S Rose}, \bibinfo{person}{Youngok Pino},
  \bibinfo{person}{Ozgur Sinanoglu}, {and} \bibinfo{person}{Ramesh Karri}.}
  \bibinfo{year}{2015}\natexlab{}.
\newblock \showarticletitle{{Fault analysis-based logic encryption}}.
\newblock \bibinfo{journal}{\emph{IEEE Transactions on computers}}
  \bibinfo{volume}{64}, \bibinfo{number}{2} (\bibinfo{year}{2015}),
  \bibinfo{pages}{410--424}.
\newblock


\bibitem[\protect\citeauthoryear{Rajendran, Sinanoglu, and Karri}{Rajendran
  et~al\mbox{.}}{2013b}]%
        {rajendran2013split}
\bibfield{author}{\bibinfo{person}{J.~J.~V. Rajendran}, \bibinfo{person}{Ozgur
  Sinanoglu}, {and} \bibinfo{person}{Ramesh Karri}.}
  \bibinfo{year}{2013}\natexlab{b}.
\newblock \showarticletitle{{Is Split Manufacturing Secure?}}. In
  \bibinfo{booktitle}{\emph{Proc. Conf. Design, Automation and Test in Europe
  (DATE)}}. \bibinfo{pages}{1259--1264}.
\newblock


\bibitem[\protect\citeauthoryear{Ramdas, Saeed, and Sinanoglu}{Ramdas
  et~al\mbox{.}}{2014}]%
        {ramdas2014slack}
\bibfield{author}{\bibinfo{person}{Abishek Ramdas},
  \bibinfo{person}{Samah~Mohamed Saeed}, {and} \bibinfo{person}{Ozgur
  Sinanoglu}.} \bibinfo{year}{2014}\natexlab{}.
\newblock \showarticletitle{Slack removal for enhanced reliability and trust}.
  In \bibinfo{booktitle}{\emph{{Int. Conference on Design \& Technology of
  Integrated Systems in Nanoscale Era (DTIS)}}}. \bibinfo{pages}{1--4}.
\newblock


\bibitem[\protect\citeauthoryear{Roy, Koushanfar, and Markov}{Roy
  et~al\mbox{.}}{2008}]%
        {roy2008epic}
\bibfield{author}{\bibinfo{person}{Jarrod~A Roy}, \bibinfo{person}{Farinaz
  Koushanfar}, {and} \bibinfo{person}{Igor~L Markov}.}
  \bibinfo{year}{2008}\natexlab{}.
\newblock \showarticletitle{{EPIC: Ending piracy of integrated circuits}}. In
  \bibinfo{booktitle}{\emph{Proceedings of the conference on Design, automation
  and test in Europe}}. \bibinfo{pages}{1069--1074}.
\newblock


\bibitem[\protect\citeauthoryear{Salmani, Tehranipoor, and Plusquellic}{Salmani
  et~al\mbox{.}}{2012}]%
        {salmani2012novel}
\bibfield{author}{\bibinfo{person}{Hassan Salmani}, \bibinfo{person}{Mohammad
  Tehranipoor}, {and} \bibinfo{person}{Jim Plusquellic}.}
  \bibinfo{year}{2012}\natexlab{}.
\newblock \showarticletitle{{{A Novel Technique for Improving Hardware Trojan
  Detection and Reducing Trojan Activation Time}}}.
\newblock \bibinfo{journal}{\emph{IEEE Trans. Very Large Scale Integration
  Sys.}} (\bibinfo{year}{2012}), \bibinfo{pages}{112--125}.
\newblock


\bibitem[\protect\citeauthoryear{Sengupta and Mohanty}{Sengupta and
  Mohanty}{2018}]%
        {sengupta2018functional}
\bibfield{author}{\bibinfo{person}{Anirban Sengupta} {and}
  \bibinfo{person}{Saraju~P Mohanty}.} \bibinfo{year}{2018}\natexlab{}.
\newblock \showarticletitle{{Functional obfuscation of DSP cores using robust
  logic locking and encryption}}. In \bibinfo{booktitle}{\emph{Computer Society
  Annual Symposium on VLSI (ISVLSI)}}. \bibinfo{pages}{709--713}.
\newblock


\bibitem[\protect\citeauthoryear{Shamsi, Li, Meade, Zhao, Pan, and Jin}{Shamsi
  et~al\mbox{.}}{2017}]%
        {shamsi2017appsat}
\bibfield{author}{\bibinfo{person}{Kaveh Shamsi}, \bibinfo{person}{Meng Li},
  \bibinfo{person}{Travis Meade}, \bibinfo{person}{Zheng Zhao},
  \bibinfo{person}{David~Z Pan}, {and} \bibinfo{person}{Yier Jin}.}
  \bibinfo{year}{2017}\natexlab{}.
\newblock \showarticletitle{{AppSAT: Approximately deobfuscating integrated
  circuits}}. In \bibinfo{booktitle}{\emph{Int. Symposium on Hardware Oriented
  Security and Trust (HOST)}}. \bibinfo{pages}{95--100}.
\newblock


\bibitem[\protect\citeauthoryear{Shen, Asadizanjani, Tehranipoor, and
  Forte}{Shen et~al\mbox{.}}{2018}]%
        {shen2018nanopyramid}
\bibfield{author}{\bibinfo{person}{Haoting Shen}, \bibinfo{person}{Navid
  Asadizanjani}, \bibinfo{person}{Mark Tehranipoor}, {and}
  \bibinfo{person}{Domenic Forte}.} \bibinfo{year}{2018}\natexlab{}.
\newblock \showarticletitle{{Nanopyramid: An Optical Scrambler Against Backside
  Probing Attacks}}. In \bibinfo{booktitle}{\emph{Proc. Int. Symposium for
  Testing and Failure Analysis(ISTFA)}}. \bibinfo{pages}{280}.
\newblock


\bibitem[\protect\citeauthoryear{Shen and Zhou}{Shen and Zhou}{2017}]%
        {shen2017double}
\bibfield{author}{\bibinfo{person}{Yuanqi Shen} {and} \bibinfo{person}{Hai
  Zhou}.} \bibinfo{year}{2017}\natexlab{}.
\newblock \showarticletitle{{Double dip: Re-evaluating security of logic
  encryption algorithms}}. In \bibinfo{booktitle}{\emph{Proceedings of the
  Great Lakes Symposium on VLSI}}. \bibinfo{pages}{179--184}.
\newblock


\bibitem[\protect\citeauthoryear{Shi, Xiao, Forte, and Tehranipoor}{Shi
  et~al\mbox{.}}{2017}]%
        {shi2017obfuscated}
\bibfield{author}{\bibinfo{person}{Qihang Shi}, \bibinfo{person}{Kan Xiao},
  \bibinfo{person}{Domenic Forte}, {and} \bibinfo{person}{Mark~M Tehranipoor}.}
  \bibinfo{year}{2017}\natexlab{}.
\newblock \showarticletitle{Obfuscated built-in self-authentication}.
\newblock In \bibinfo{booktitle}{\emph{Hardware Protection through
  Obfuscation}}. \bibinfo{publisher}{Springer}, \bibinfo{pages}{263--289}.
\newblock


\bibitem[\protect\citeauthoryear{Sinanoglu, Karimi, Rajendran, Karri, Jin,
  Huang, and Makris}{Sinanoglu et~al\mbox{.}}{2013}]%
        {Ozgur2013}
\bibfield{author}{\bibinfo{person}{O. Sinanoglu}, \bibinfo{person}{N. Karimi},
  \bibinfo{person}{J. Rajendran}, \bibinfo{person}{R. Karri},
  \bibinfo{person}{Y. Jin}, \bibinfo{person}{K. Huang}, {and}
  \bibinfo{person}{Y. Makris}.} \bibinfo{year}{2013}\natexlab{}.
\newblock \showarticletitle{{{Reconciling the IC Test and Security
  Dichotomy}}}. In \bibinfo{booktitle}{\emph{Proc. of IEEE European Test
  Symp.}}
\newblock


\bibitem[\protect\citeauthoryear{Sirone and Subramanyan}{Sirone and
  Subramanyan}{2018}]%
        {sirone2018functional}
\bibfield{author}{\bibinfo{person}{Deepak Sirone} {and} \bibinfo{person}{Pramod
  Subramanyan}.} \bibinfo{year}{2018}\natexlab{}.
\newblock \showarticletitle{{Functional Analysis Attacks on Logic Locking}}.
\newblock \bibinfo{journal}{\emph{arXiv preprint arXiv:1811.12088}}
  (\bibinfo{year}{2018}).
\newblock


\bibitem[\protect\citeauthoryear{Sturton, Hicks, Wagner, and King}{Sturton
  et~al\mbox{.}}{2011}]%
        {sturton2011defeating}
\bibfield{author}{\bibinfo{person}{Cynthia Sturton}, \bibinfo{person}{Matthew
  Hicks}, \bibinfo{person}{David Wagner}, {and} \bibinfo{person}{Samuel~T
  King}.} \bibinfo{year}{2011}\natexlab{}.
\newblock \showarticletitle{Defeating UCI: Building stealthy and malicious
  hardware}. In \bibinfo{booktitle}{\emph{2011 IEEE Symposium on Security and
  Privacy}}. IEEE, \bibinfo{pages}{64--77}.
\newblock


\bibitem[\protect\citeauthoryear{Subramani, Antonopoulos, Abotabl, Nosratinia,
  and Makris}{Subramani et~al\mbox{.}}{2017}]%
        {subramani2017ace}
\bibfield{author}{\bibinfo{person}{Kiruba~Sankaran Subramani},
  \bibinfo{person}{Angelos Antonopoulos}, \bibinfo{person}{Ahmed~Attia
  Abotabl}, \bibinfo{person}{Aria Nosratinia}, {and} \bibinfo{person}{Yiorgos
  Makris}.} \bibinfo{year}{2017}\natexlab{}.
\newblock \showarticletitle{ACE: Adaptive channel estimation for detecting
  analog/RF trojans in WLAN transceivers}. In \bibinfo{booktitle}{\emph{2017
  IEEE/ACM International Conference on Computer-Aided Design (ICCAD)}}. IEEE,
  \bibinfo{pages}{722--727}.
\newblock


\bibitem[\protect\citeauthoryear{Subramanyan, Ray, and Malik}{Subramanyan
  et~al\mbox{.}}{2015}]%
        {subramanyan2015evaluating}
\bibfield{author}{\bibinfo{person}{Pramod Subramanyan}, \bibinfo{person}{Sayak
  Ray}, {and} \bibinfo{person}{Sharad Malik}.} \bibinfo{year}{2015}\natexlab{}.
\newblock \showarticletitle{Evaluating the security of logic encryption
  algorithms}. In \bibinfo{booktitle}{\emph{IEEE International Symposium on
  Hardware Oriented Security and Trust (HOST)}}. \bibinfo{pages}{137--143}.
\newblock


\bibitem[\protect\citeauthoryear{{Synopsys Inc., Mountain View, CA,
  USA}}{{Synopsys Inc., Mountain View, CA, USA}}{2017}]%
        {SynopsysTetraMAX}
\bibfield{author}{\bibinfo{person}{{Synopsys Inc., Mountain View, CA, USA}}.}
  \bibinfo{year}{2017}\natexlab{}.
\newblock \bibinfo{title}{{TetraMAX ATPG: Automatic Test Pattern Generation}}.
\newblock
\newblock


\bibitem[\protect\citeauthoryear{Tehranipoor and Koushanfar}{Tehranipoor and
  Koushanfar}{2010}]%
        {tehranipoor2010survey}
\bibfield{author}{\bibinfo{person}{Mohammad Tehranipoor} {and}
  \bibinfo{person}{Farinaz Koushanfar}.} \bibinfo{year}{2010}\natexlab{}.
\newblock \showarticletitle{{{A Survey of Hardware Trojan Taxonomy and
  Detection}}}.
\newblock \bibinfo{journal}{\emph{IEEE Design \& Test of Computers}}
  \bibinfo{volume}{27}, \bibinfo{number}{1} (\bibinfo{year}{2010}).
\newblock


\bibitem[\protect\citeauthoryear{Tehranipoor, Salmani, Zhang, Wang, Karri,
  Rajendran, and Rosenfeld}{Tehranipoor et~al\mbox{.}}{2011}]%
        {tehranipoor2011trustworthy}
\bibfield{author}{\bibinfo{person}{Mohammad Tehranipoor},
  \bibinfo{person}{Hassan Salmani}, \bibinfo{person}{Xuehui Zhang},
  \bibinfo{person}{Michel Wang}, \bibinfo{person}{Ramesh Karri},
  \bibinfo{person}{Jeyavijayan Rajendran}, {and} \bibinfo{person}{Kurt
  Rosenfeld}.} \bibinfo{year}{2011}\natexlab{}.
\newblock \showarticletitle{{{Trustworthy Hardware: Trojan Detection and
  Design-for-Trust Challenges}}}.
\newblock \bibinfo{journal}{\emph{Computer}} (\bibinfo{year}{2011}).
\newblock


\bibitem[\protect\citeauthoryear{Tehranipoor and Wang}{Tehranipoor and
  Wang}{2011}]%
        {tehranipoor2011introduction}
\bibfield{author}{\bibinfo{person}{Mohammad Tehranipoor} {and}
  \bibinfo{person}{Cliff Wang}.} \bibinfo{year}{2011}\natexlab{}.
\newblock \bibinfo{booktitle}{\emph{Introduction to hardware security and
  trust}}.
\newblock \bibinfo{publisher}{Springer Science \& Business Media}.
\newblock


\bibitem[\protect\citeauthoryear{Tehranipoor, Guin, and Forte}{Tehranipoor
  et~al\mbox{.}}{2015}]%
        {tehranipoor2015counterfeit}
\bibfield{author}{\bibinfo{person}{Mark~Mohammad Tehranipoor},
  \bibinfo{person}{Ujjwal Guin}, {and} \bibinfo{person}{Domenic Forte}.}
  \bibinfo{year}{2015}\natexlab{}.
\newblock \showarticletitle{Counterfeit integrated circuits}.
\newblock In \bibinfo{booktitle}{\emph{Counterfeit Integrated Circuits}}.
  \bibinfo{publisher}{Springer}, \bibinfo{pages}{15--36}.
\newblock


\bibitem[\protect\citeauthoryear{Torrance and James}{Torrance and
  James}{2009}]%
        {torrance2009state}
\bibfield{author}{\bibinfo{person}{Randy Torrance} {and} \bibinfo{person}{Dick
  James}.} \bibinfo{year}{2009}\natexlab{}.
\newblock \showarticletitle{{The state-of-the-art in IC reverse engineering}}.
  In \bibinfo{booktitle}{\emph{International Workshop on Cryptographic Hardware
  and Embedded Systems}}. \bibinfo{pages}{363--381}.
\newblock


\bibitem[\protect\citeauthoryear{Vaidyanathan, Das, and Pileggi}{Vaidyanathan
  et~al\mbox{.}}{2014a}]%
        {vaidyanathan2014detecting}
\bibfield{author}{\bibinfo{person}{Kaushik Vaidyanathan},
  \bibinfo{person}{Bishnu~P Das}, {and} \bibinfo{person}{Larry Pileggi}.}
  \bibinfo{year}{2014}\natexlab{a}.
\newblock \showarticletitle{{{Detecting Reliability Attacks During Split
  Fabrication Using Test-Only BEOL Stack}}}. In \bibinfo{booktitle}{\emph{Proc.
  of Design Automation Conf.}} \bibinfo{pages}{1--6}.
\newblock


\bibitem[\protect\citeauthoryear{Vaidyanathan, Liu, Sumbul, Zhu, Franchetti,
  and Pileggi}{Vaidyanathan et~al\mbox{.}}{2014b}]%
        {vaidyanathan2014efficient}
\bibfield{author}{\bibinfo{person}{Kaushik Vaidyanathan},
  \bibinfo{person}{Renzhi Liu}, \bibinfo{person}{Ekin Sumbul},
  \bibinfo{person}{Qiuling Zhu}, \bibinfo{person}{Franz Franchetti}, {and}
  \bibinfo{person}{Larry Pileggi}.} \bibinfo{year}{2014}\natexlab{b}.
\newblock \showarticletitle{{Efficient and secure intellectual property (IP)
  design with split fabrication}}. In \bibinfo{booktitle}{\emph{IEEE
  International Symposium on Hardware-Oriented Security and Trust (HOST)}}.
  \bibinfo{pages}{13--18}.
\newblock


\bibitem[\protect\citeauthoryear{Vashistha, Lu, Shi, Rahman, Shen, Woodard,
  Asadizanjani, and Tehranipoor}{Vashistha et~al\mbox{.}}{2018}]%
        {vashistha2018trojan}
\bibfield{author}{\bibinfo{person}{Nidish Vashistha}, \bibinfo{person}{Hangwei
  Lu}, \bibinfo{person}{Qihang Shi}, \bibinfo{person}{M~Tanjidur Rahman},
  \bibinfo{person}{Haoting Shen}, \bibinfo{person}{Damon~L Woodard},
  \bibinfo{person}{Navid Asadizanjani}, {and} \bibinfo{person}{Mark
  Tehranipoor}.} \bibinfo{year}{2018}\natexlab{}.
\newblock \showarticletitle{Trojan Scanner: Detecting Hardware Trojans with
  Rapid SEM Imaging combined with Image Processing and Machine Learning}. In
  \bibinfo{booktitle}{\emph{Proc. Int. Symposium for Testing and Failure
  Analysis}}. \bibinfo{pages}{256}.
\newblock


\bibitem[\protect\citeauthoryear{Waksman, Suozzo, and Sethumadhavan}{Waksman
  et~al\mbox{.}}{2013}]%
        {waksman2013fanci}
\bibfield{author}{\bibinfo{person}{Adam Waksman}, \bibinfo{person}{Matthew
  Suozzo}, {and} \bibinfo{person}{Simha Sethumadhavan}.}
  \bibinfo{year}{2013}\natexlab{}.
\newblock \showarticletitle{{{FANCI: Identification of Stealthy Malicious Logic
  Using Boolean Functional Analysis}}}. In \bibinfo{booktitle}{\emph{Proc. ACM
  SIGSAC Conf. on Computer \& Communications Security}}.
  \bibinfo{pages}{697--708}.
\newblock


\bibitem[\protect\citeauthoryear{Wang, Forte, Tehranipoor, and Shi}{Wang
  et~al\mbox{.}}{2017b}]%
        {wang2017probing}
\bibfield{author}{\bibinfo{person}{Huanyu Wang}, \bibinfo{person}{Domenic
  Forte}, \bibinfo{person}{Mark~M Tehranipoor}, {and} \bibinfo{person}{Qihang
  Shi}.} \bibinfo{year}{2017}\natexlab{b}.
\newblock \showarticletitle{{Probing attacks on integrated circuits: Challenges
  and research opportunities}}.
\newblock \bibinfo{journal}{\emph{IEEE Design \& Test}} \bibinfo{volume}{34},
  \bibinfo{number}{5} (\bibinfo{year}{2017}), \bibinfo{pages}{63--71}.
\newblock


\bibitem[\protect\citeauthoryear{Wang, Shi, Forte, and Tehranipoor}{Wang
  et~al\mbox{.}}{2019}]%
        {wang2019probing}
\bibfield{author}{\bibinfo{person}{Huanyu Wang}, \bibinfo{person}{Qihang Shi},
  \bibinfo{person}{Domenic Forte}, {and} \bibinfo{person}{Mark~M Tehranipoor}.}
  \bibinfo{year}{2019}\natexlab{}.
\newblock \showarticletitle{{Probing Assessment Framework and Evaluation of
  Antiprobing Solutions}}.
\newblock \bibinfo{journal}{\emph{Transactions on Very Large Scale Integration
  (VLSI) Systems}} \bibinfo{volume}{27}, \bibinfo{number}{6}
  (\bibinfo{year}{2019}), \bibinfo{pages}{1239--1252}.
\newblock


\bibitem[\protect\citeauthoryear{Wang, Narasimhan, Krishna, Mal-Sarkar, and
  Bhunia}{Wang et~al\mbox{.}}{2011}]%
        {wang2011sequential}
\bibfield{author}{\bibinfo{person}{Xinmu Wang}, \bibinfo{person}{Seetharam
  Narasimhan}, \bibinfo{person}{Aswin Krishna}, \bibinfo{person}{Tatini
  Mal-Sarkar}, {and} \bibinfo{person}{Swarup Bhunia}.}
  \bibinfo{year}{2011}\natexlab{}.
\newblock \showarticletitle{{Sequential hardware trojan: Side-channel aware
  design and placement}}. In \bibinfo{booktitle}{\emph{International Conference
  on Computer Design (ICCD)}}. \bibinfo{pages}{297--300}.
\newblock


\bibitem[\protect\citeauthoryear{Wang, Cao, Hu, and Rajendran}{Wang
  et~al\mbox{.}}{2017a}]%
        {wang2017front}
\bibfield{author}{\bibinfo{person}{Yujie Wang}, \bibinfo{person}{Tri Cao},
  \bibinfo{person}{Jiang Hu}, {and} \bibinfo{person}{Jeyavijayan Rajendran}.}
  \bibinfo{year}{2017}\natexlab{a}.
\newblock \showarticletitle{Front-end-of-line attacks in split manufacturing}.
  In \bibinfo{booktitle}{\emph{IEEE/ACM International Conference on
  Computer-Aided Design (ICCAD)}}. \bibinfo{pages}{1--8}.
\newblock


\bibitem[\protect\citeauthoryear{Wang, Chen, Hu, and Rajendran}{Wang
  et~al\mbox{.}}{2016}]%
        {wang2016cat}
\bibfield{author}{\bibinfo{person}{Yujie Wang}, \bibinfo{person}{Pu Chen},
  \bibinfo{person}{Jiang Hu}, {and} \bibinfo{person}{Jeyavijayan~JV
  Rajendran}.} \bibinfo{year}{2016}\natexlab{}.
\newblock \showarticletitle{{{The Cat and Mouse in Split Manufacturing}}}. In
  \bibinfo{booktitle}{\emph{Proceedings of Design Automation Conference}}.
  \bibinfo{pages}{1--6}.
\newblock


\bibitem[\protect\citeauthoryear{Wei, Meguerdichian, and M.Potkonjak}{Wei
  et~al\mbox{.}}{2011}]%
        {Wei_2011}
\bibfield{author}{\bibinfo{person}{S. Wei}, \bibinfo{person}{S. Meguerdichian},
  {and} \bibinfo{person}{M.Potkonjak}.} \bibinfo{year}{2011}\natexlab{}.
\newblock \showarticletitle{{{Malicious Circuitry Detection Using Thermal
  Conditioning}}}.
\newblock \bibinfo{journal}{\emph{Trans. of Information, Forensics and
  Security}} \bibinfo{volume}{6}, \bibinfo{number}{3} (\bibinfo{year}{2011}),
  \bibinfo{pages}{1136–--1145}.
\newblock


\bibitem[\protect\citeauthoryear{Xiao and Tehranipoor}{Xiao and
  Tehranipoor}{2013}]%
        {xiao2013bisa}
\bibfield{author}{\bibinfo{person}{Kan Xiao} {and} \bibinfo{person}{Mohammed
  Tehranipoor}.} \bibinfo{year}{2013}\natexlab{}.
\newblock \showarticletitle{{{BISA: Built-In Self-Authentication for Preventing
  Hardware Trojan Insertion}}}. In \bibinfo{booktitle}{\emph{Proc. IEEE Int.
  Symp. Hardware-Oriented Security and Trust}}. \bibinfo{pages}{45--50}.
\newblock


\bibitem[\protect\citeauthoryear{Xie and Srivastava}{Xie and
  Srivastava}{2016}]%
        {xie2016mitigating}
\bibfield{author}{\bibinfo{person}{Yang Xie} {and} \bibinfo{person}{Ankur
  Srivastava}.} \bibinfo{year}{2016}\natexlab{}.
\newblock \showarticletitle{{Mitigating SAT attack on logic locking}}. In
  \bibinfo{booktitle}{\emph{International Conference on Cryptographic Hardware
  and Embedded Systems}}. \bibinfo{pages}{127--146}.
\newblock


\bibitem[\protect\citeauthoryear{Xie and Srivastava}{Xie and
  Srivastava}{2019}]%
        {xie2019anti}
\bibfield{author}{\bibinfo{person}{Yang Xie} {and} \bibinfo{person}{Ankur
  Srivastava}.} \bibinfo{year}{2019}\natexlab{}.
\newblock \showarticletitle{{Anti-sat: Mitigating sat attack on logic
  locking}}.
\newblock \bibinfo{journal}{\emph{IEEE Transactions on Computer-Aided Design of
  Integrated Circuits and Systems}} \bibinfo{volume}{38}, \bibinfo{number}{2}
  (\bibinfo{year}{2019}), \bibinfo{pages}{199--207}.
\newblock


\bibitem[\protect\citeauthoryear{Xu, Feng, Rajendran, and Hu}{Xu
  et~al\mbox{.}}{2019}]%
        {xu2019layout}
\bibfield{author}{\bibinfo{person}{Wenbin Xu}, \bibinfo{person}{Lang Feng},
  \bibinfo{person}{Jeyavijayan~JV Rajendran}, {and} \bibinfo{person}{Jiang
  Hu}.} \bibinfo{year}{2019}\natexlab{}.
\newblock \showarticletitle{Layout recognition attacks on split manufacturing}.
  In \bibinfo{booktitle}{\emph{Proceedings of Asia and South Pacific Design
  Automation Conference}}. \bibinfo{pages}{45--50}.
\newblock


\bibitem[\protect\citeauthoryear{Xu, Shakya, Tehranipoor, and Forte}{Xu
  et~al\mbox{.}}{2017}]%
        {xu2017novel}
\bibfield{author}{\bibinfo{person}{Xiaolin Xu}, \bibinfo{person}{Bicky Shakya},
  \bibinfo{person}{Mark~M Tehranipoor}, {and} \bibinfo{person}{Domenic Forte}.}
  \bibinfo{year}{2017}\natexlab{}.
\newblock \showarticletitle{{Novel bypass attack and BDD-based tradeoff
  analysis against all known logic locking attacks}}. In
  \bibinfo{booktitle}{\emph{Int. Conf. on Cryptographic Hardware and Embedded
  Systems}}.
\newblock


\bibitem[\protect\citeauthoryear{Yang, Hicks, Dong, Austin, and Sylvester}{Yang
  et~al\mbox{.}}{2016}]%
        {yang2016a2}
\bibfield{author}{\bibinfo{person}{Kaiyuan Yang}, \bibinfo{person}{Matthew
  Hicks}, \bibinfo{person}{Qing Dong}, \bibinfo{person}{Todd Austin}, {and}
  \bibinfo{person}{Dennis Sylvester}.} \bibinfo{year}{2016}\natexlab{}.
\newblock \showarticletitle{{A2: Analog malicious hardware}}. In
  \bibinfo{booktitle}{\emph{IEEE symposium on security and privacy (SP)}}.
  \bibinfo{pages}{18--37}.
\newblock


\bibitem[\protect\citeauthoryear{Yasin, Mazumdar, Rajendran, and
  Sinanoglu}{Yasin et~al\mbox{.}}{2016}]%
        {yasin2016sarlock}
\bibfield{author}{\bibinfo{person}{Muhammad Yasin}, \bibinfo{person}{Bodhisatwa
  Mazumdar}, \bibinfo{person}{Jeyavijayan~JV Rajendran}, {and}
  \bibinfo{person}{Ozgur Sinanoglu}.} \bibinfo{year}{2016}\natexlab{}.
\newblock \showarticletitle{{SARLock: SAT attack resistant logic locking}}. In
  \bibinfo{booktitle}{\emph{IEEE International Symposium on Hardware Oriented
  Security and Trust (HOST)}}. \bibinfo{pages}{236--241}.
\newblock


\bibitem[\protect\citeauthoryear{Yasin, Rajendran, Sinanoglu, and Karri}{Yasin
  et~al\mbox{.}}{2015}]%
        {yasin2015improving}
\bibfield{author}{\bibinfo{person}{Muhammad Yasin},
  \bibinfo{person}{Jeyavijayan~JV Rajendran}, \bibinfo{person}{Ozgur
  Sinanoglu}, {and} \bibinfo{person}{Ramesh Karri}.}
  \bibinfo{year}{2015}\natexlab{}.
\newblock \showarticletitle{{On improving the security of logic locking}}.
\newblock \bibinfo{journal}{\emph{Transactions on Computer-Aided Design of
  Integrated Circuits and Systems}} (\bibinfo{year}{2015}),
  \bibinfo{pages}{1411--1424}.
\newblock


\bibitem[\protect\citeauthoryear{Yasin, Sengupta, Nabeel, Ashraf, Rajendran,
  and Sinanoglu}{Yasin et~al\mbox{.}}{2017a}]%
        {yasin2017provably}
\bibfield{author}{\bibinfo{person}{Muhammad Yasin}, \bibinfo{person}{Abhrajit
  Sengupta}, \bibinfo{person}{Mohammed~Thari Nabeel}, \bibinfo{person}{Mohammed
  Ashraf}, \bibinfo{person}{Jeyavijayan~JV Rajendran}, {and}
  \bibinfo{person}{Ozgur Sinanoglu}.} \bibinfo{year}{2017}\natexlab{a}.
\newblock \showarticletitle{{Provably-secure logic locking: From theory to
  practice}}. In \bibinfo{booktitle}{\emph{Proceedings of ACM SIGSAC Conference
  on Computer and Communications Security}}. \bibinfo{pages}{1601--1618}.
\newblock


\bibitem[\protect\citeauthoryear{Yasin, Sengupta, Schafer, Makris, Sinanoglu,
  and Rajendran}{Yasin et~al\mbox{.}}{2017b}]%
        {yasin2017lock}
\bibfield{author}{\bibinfo{person}{Muhammad Yasin}, \bibinfo{person}{Abhrajit
  Sengupta}, \bibinfo{person}{Benjamin~Carrion Schafer},
  \bibinfo{person}{Yiorgos Makris}, \bibinfo{person}{Ozgur Sinanoglu}, {and}
  \bibinfo{person}{Jeyavijayan~JV Rajendran}.}
  \bibinfo{year}{2017}\natexlab{b}.
\newblock \showarticletitle{{What to lock?: Functional and parametric
  locking}}. In \bibinfo{booktitle}{\emph{Proc. of Great Lakes Symposium on
  VLSI}}. \bibinfo{pages}{351--356}.
\newblock


\bibitem[\protect\citeauthoryear{Yasin and Sinanoglu}{Yasin and
  Sinanoglu}{2015}]%
        {yasin2015transforming}
\bibfield{author}{\bibinfo{person}{Muhammad Yasin} {and} \bibinfo{person}{Ozgur
  Sinanoglu}.} \bibinfo{year}{2015}\natexlab{}.
\newblock \showarticletitle{{Transforming between logic locking and IC
  camouflaging}}. In \bibinfo{booktitle}{\emph{International Design \& Test
  Symposium (IDT)}}. \bibinfo{pages}{1--4}.
\newblock


\bibitem[\protect\citeauthoryear{Yu, Dofe, and Zhang}{Yu et~al\mbox{.}}{2017}]%
        {yu2017exploiting}
\bibfield{author}{\bibinfo{person}{Qiaoyan Yu}, \bibinfo{person}{Jaya Dofe},
  {and} \bibinfo{person}{Zhiming Zhang}.} \bibinfo{year}{2017}\natexlab{}.
\newblock \showarticletitle{{Exploiting hardware obfuscation methods to prevent
  and detect hardware trojans}}. In \bibinfo{booktitle}{\emph{International
  Midwest Symposium on Circuits and Systems (MWSCAS)}}.
  \bibinfo{pages}{819--822}.
\newblock


\bibitem[\protect\citeauthoryear{Yu, Dofe, Zhang, and Kramer}{Yu
  et~al\mbox{.}}{2018}]%
        {yu2018hardware}
\bibfield{author}{\bibinfo{person}{Qiaoyan Yu}, \bibinfo{person}{Jaya Dofe},
  \bibinfo{person}{Zhiming Zhang}, {and} \bibinfo{person}{Sean Kramer}.}
  \bibinfo{year}{2018}\natexlab{}.
\newblock \showarticletitle{{Hardware Obfuscation Methods for Hardware Trojan
  Prevention and Detection}}.
\newblock In \bibinfo{booktitle}{\emph{The Hardware Trojan War}}.
  \bibinfo{publisher}{Springer}, \bibinfo{pages}{291--325}.
\newblock


\bibitem[\protect\citeauthoryear{Zhang, Wang, Rahman, and Tehranipoor}{Zhang
  et~al\mbox{.}}{2018}]%
        {zhang2018chip}
\bibfield{author}{\bibinfo{person}{Dongrong Zhang}, \bibinfo{person}{Xiaoxiao
  Wang}, \bibinfo{person}{Md~Tauhidur Rahman}, {and} \bibinfo{person}{Mark
  Tehranipoor}.} \bibinfo{year}{2018}\natexlab{}.
\newblock \showarticletitle{{An On-Chip Dynamically Obfuscated Wrapper for
  Protecting Supply Chain Against IP and IC Piracies}}.
\newblock \bibinfo{journal}{\emph{IEEE Transactions on Very Large Scale
  Integration (VLSI) Systems}} \bibinfo{volume}{26}, \bibinfo{number}{11}
  (\bibinfo{year}{2018}), \bibinfo{pages}{2456--2469}.
\newblock


\bibitem[\protect\citeauthoryear{Zhang, Cui, Zhou, and Guin}{Zhang
  et~al\mbox{.}}{2019}]%
        {zhang2019tga}
\bibfield{author}{\bibinfo{person}{Yuqiao Zhang}, \bibinfo{person}{Pinchen
  Cui}, \bibinfo{person}{Ziqi Zhou}, {and} \bibinfo{person}{Ujjwal Guin}.}
  \bibinfo{year}{2019}\natexlab{}.
\newblock \showarticletitle{{TGA: An Oracle-less and Topology-Guided Attack on
  Logic Locking}}. In \bibinfo{booktitle}{\emph{Proc. of the ACM Workshop on
  Attacks and Solutions in Hardware Security Workshop (ASHES)}}.
  \bibinfo{pages}{75--83}.
\newblock


\bibitem[\protect\citeauthoryear{Zhou, Zhang, Thambipillai, and Teo}{Zhou
  et~al\mbox{.}}{2014}]%
        {zhou2014low}
\bibfield{author}{\bibinfo{person}{Bin Zhou}, \bibinfo{person}{Wei Zhang},
  \bibinfo{person}{Srikanthan Thambipillai}, {and} \bibinfo{person}{JKJ Teo}.}
  \bibinfo{year}{2014}\natexlab{}.
\newblock \showarticletitle{{A low cost acceleration method for hardware Trojan
  detection based on fan-out cone analysis}}. In
  \bibinfo{booktitle}{\emph{Proceedings of International Conference on
  Hardware/Software Codesign and System Synthesis}}. \bibinfo{pages}{28}.
\newblock


\bibitem[\protect\citeauthoryear{Zhou, Guin, and Agrawal}{Zhou
  et~al\mbox{.}}{2018}]%
        {zhou2018modeling}
\bibfield{author}{\bibinfo{person}{Ziqi Zhou}, \bibinfo{person}{Ujjwal Guin},
  {and} \bibinfo{person}{Vishwani~D Agrawal}.} \bibinfo{year}{2018}\natexlab{}.
\newblock \showarticletitle{Modeling and test generation for combinational
  hardware Trojans}. In \bibinfo{booktitle}{\emph{VLSI Test Symposium (VTS)}}.
  \bibinfo{pages}{1--6}.
\newblock


\end{thebibliography}

%\vspace{-20px}

\end{document}